\documentclass[iop]{emulateapj} 
\slugcomment{{\sc Accepted to ApJ:} July 28, 2015} 
\usepackage{natbib,amsmath,booktabs,apjfonts,threeparttable,multirow}
\usepackage{natbib}
\usepackage{graphicx}
\usepackage{tabularx}
\usepackage{epstopdf}
\usepackage{color}

\begin{document}

\title{Multi-wavelength analysis of the Galactic supernova remnant MSH~11$-$6{\sl 1}A}

\author{Katie Auchettl\altaffilmark{1,2}, 
        Patrick Slane\altaffilmark{1}, Daniel Castro\altaffilmark{3}, Adam R. Foster\altaffilmark{1}, Randall K. Smith\altaffilmark{1}}

\altaffiltext{1}{Harvard-Smithsonian Center for Astrophysics, 60 Garden Street, Cambridge, MA 02138, USA}
\altaffiltext{2}{School of Physics and Astronomy, Monash University, Melbourne, Victoria 3800, Australia}
\altaffiltext{3}{MIT-Kavli Center for Astrophysics and Space Research, 77 Massachusetts Avenue, Cambridge, MA 02139, USA}

\begin{abstract}
Due to its centrally bright X-ray morphology and limb brightened radio profile, MSH~11$-$6{\sl 1}A (G290.1-0.8) is classified as a mixed morphology supernova remnant (SNR). H$\textsc{i}$ and CO observations determined that the SNR is interacting with molecular clouds found toward the north and southwest regions of the remnant. 
In this paper we report on the detection of $\gamma$-ray emission coincident with MSH~11$-$6{\sl 1}A, using 70 months of data from the Large Area Telescope on board the \textit{Fermi Gamma-ray Space Telescope}. To investigate the origin of this emission, we perform broadband modelling of its non-thermal emission considering both leptonic and hadronic cases and concluding that the $\gamma$-ray emission is most likely hadronic in nature. Additionally we present our analysis of a 111 ks archival $Suzaku$ observation of this remnant. Our investigation shows that the X-ray emission from MSH~11$-$6{\sl 1}A arises from shock-heated ejecta with the bulk of the X-ray emission arising from a recombining plasma, while the emission towards the east arises from an ionising plasma.
\end{abstract}


\section{INTRODUCTION}
Since the launch of the Large Area Telescope (LAT) onboard the \textit{Fermi Gamma-ray Space Telescope}, its improved sensitivity and resolution in the MeV-GeV energy range has lead to a number of supernova remnants (SNRs) being detected in GeV $\gamma$-rays. The shock-front of an SNR is expected to be able to accelerate cosmic rays (CRs) efficiently, producing non-thermal X-ray and $\gamma$-ray emission. As $\gamma$-rays can arise from leptonic processes such as inverse Compton (IC) scattering and non-thermal bremsstrahlung from high-energy electrons, or from hadronic emission arising from the decay of a neutral pion (produced in a proton-proton interaction) into two photons, a means of distinguishing between these two mechanisms is crucial for our understanding of the origin of this observed emission. Thermal and non-thermal emission from SNRs have provided increasing support in favour of CRs being accelerated at the shock front of the remnant (e.g. Tycho: \citealp{2005ApJ...634..376W}; RX J1713.7-3946: \citealp{2007Natur.449..576U}; W44, MSH 17-39 and G337.7-0.1: \citealp{2013ApJ...774...36C}).  SNRs known to be interacting with dense molecular clouds (MCs) are ideal, indirect laboratories that one can use to detect and analyse $\gamma$-rays arising from accelerated protons. The interaction of the SNR's shockwave with dense molecular material is often inferred from the detection of one or more OH (1720 MHz) masers, but enhancement of excitation line ratios such as $J=2\rightarrow1/ J = 1 \rightarrow 0$, broadenings and asymmetries in molecular line features or morphological alignment of molecular features with SNR features can also allow one to determine SNR/MC interaction (see \citealp{2014SSRv..tmp...26S} and references therein).
  
Some of the first SNRs detected by the \textit{Fermi-LAT} (e.g. W44: \citealt{2010Sci...327.1103A}, \citealt{2013Sci...339..807A}; IC443: \citep{2013Sci...339..807A}; 3C391: \citealp{2010ApJ...717..372C}; and W49B: \citealp{2010ApJ...722.1303A}) are part of a unique class called Mixed-Morphology (MM) SNRs. Some of these SNRs are known to be interacting with MCs. These SNRs are characterised by their centrally peaked X-ray morphology which is thermal in nature, while their radio profiles are limb-brightened \citep{1998ApJ...503L.167R}. The evolutionary sequence leading to these unusual X-ray properties are not well understood and the morphology and characteristics of these SNRs are difficult to explain using standard SNR evolution models. There are two main models that are invoked in the literature to explain their characteristics. One possible model \citep{1991ApJ...373..543W} assumes that in the vicinity of the supernova explosion there are many small, dense, cold cloudlets. These cloudlets are small enough that they do not affect the passage of the shock, and are sufficiently dense that they are neither blown apart nor swept up. Once the shock has passed, the cloudlets slowly evaporate, filling the interior of the SNR with a relatively dense gas that emits in X-rays. Another possible scenario is that thermal conduction results in the transport of heat and material to the center of the remnant, increasing the central density of the remnant, and smoothing the temperature gradient behind the shock \citep{1999ApJ...524..179C}. 
  
The ionisation state of a thermal plasma in an SNR can be characterised by its ionisation temperature ($T_{Z}$) which describes the extent to which the ions are stripped of their electrons and its electron temperature ($T_{e}$) which describes the kinetic energy of the electrons. The thermal plasmas of SNRs have been thought to be either underionised, where $T_{Z} < T_{e}$ or in collisional ionisation equilibrium $T_{Z} = T_{e}$. Recent observations by the $Suzaku$ satellite have confirmed earlier suggestions based on $ASCA$ data \citep{2005ApJ...631..935K}, that the thermal plasma in some MM SNRs is overionised (recombining) (e.g. 3C391: \citealp{2014arXiv1409.6861S, 2014ApJ...790...65E}). Recombining plasmas have ionisation temperatures that are higher than the electron temperatures and require rapid cooling of electrons either by thermal conduction \citep{2002ApJ...572..897K}, adiabatic expansion via rarefraction and recombination \citep{1989MNRAS.236..885I} or the interaction with dense cavity walls or molecular clouds \citep{2005ApJ...630..892D}. 
  
MSH~11$-$6{\sl 1}A (G290.1-0.8) is a Galactic MM SNR that is known to be interacting with MCs. It was first discovered by \citet{1961AuJPh..14..497M} using the Sydney 3.5m cross-type radio telescope. It was first identified as an SNR by \citet{1968AuJPh..21..369K} and later classified as a shell-type SNR with a complex internal structure and ear-like protrusions towards the northwest and southeast using the Molongo Observatory Synthesis Telescope (MOST) at multiple different wavelengths \citep{1987A&A...183..118K,1989PASAu...8..187M, 1996A&AS..118..329W}. It has an angular size of 19' $\times$ 11' and a radio-continuum spectral index of $\alpha = -0.33 +/- 0.07$ \citep{2006MNRAS.369..416R}. Radio continuum observations using the Australia Telescope Compact Array (ATCA) by \citet{2006MNRAS.369..416R} showed filamentary emission with little shell structure, while the northern and southern edges of the remnant show evidence that the shock front could be interacting with a plane parallel density gradient. Using NANTEN CO images of MSH~11$-$6{\sl 1}A, \citet{2005SerAJ.170...47F} determine that the SNR is associated with a MC towards the south-west and northern rim of the remnant. \textsc{Hi} observations using the Southern Galactic Plane Survey find that the molecular cloud is found at a local standard of rest velocity of $\sim 13$ km s$^{-1}$ \citep{2005ApJS..158..178M}.
  
MSH~11$-$6{\sl 1}A was first detected in X-rays by \citet{1990ApJS...73..781S} using a 10.9 ks Einstein Observatory observation. The 0.3 - 4.5 keV Imaging Proportional Counter (IPC) image of the remnant showed that the X-ray emission is peaked towards the center of the remnant. Using a 40 ks Advanced Satellite for Cosmology and Astrophysics ($ASCA$) GIS observation, \citet{2002ApJ...564..284S} were able to determine that the central X-ray emission is thermal in nature, classifying the remnant as a MM SNR. They modelled the X-ray emission using the cloudy ISM model by \citet{1991ApJ...373..543W} and derived an intercloud medium density of $\sim 0.05 - 0.40$ cm$^{-3}$ and an age of 10 - 20 kyr. Using \textit{XMM-Newton}, \citet{2012A&A...546A..91G} analysed five regions along the axes of the remnant using an absorbed plane parallel non-equilibrium ionisation (VPSHOCK) model and found that the physical conditions across the remnant are not homogeneous, with variation in ionisation state, temperature and elemental abundances. \citet{2014arXiv1411.6809K} analysed \textit{Suzaku} data and found that in the center and in the northwest of the remnant the plasma is recombining, while everywhere else it is ionising.

The distance to MSH~11$-$6{\sl 1}A has been measured by a number of different methods. H$\textsc{i}$ measurements taken by the 64-m Parkes telescope \citep{1973ApL....15...61D,1972ApJS...24..123G} gives a lower limit of $\sim$3.5 kpc to the remnant. H$\alpha$ measurements by \citet{1996A&A...315..243R} using a Fabry-Perot interferometer implied a distance of 6.9 kpc assuming a $V_{LSR}$ of +12 km s$^{-1}$ for the SNR. Using CO measurements, \citet{2006MNRAS.369..416R}  derived a distance of 7 - 8 kpc assuming  the \citet{1993A&A...275...67B} rotation curve. \citet{2006MNRAS.369..416R} derived a distance of $7\pm1$kpc using H$\textsc{i}$ absorption measurements from ATCA, combined with data from the Southern Galactic Plane Survey, while \citet{2002ApJ...564..284S} estimated a distance of 8 - 11 kpc by modelling the thermal X-ray emission of the remnant as detected by $ASCA$. We use 7 kpc throughout this paper.
  
There are three pulsars close to the position of MSH~11$-$6{\sl 1}A. \citet{1997ApJ...485..820K} discovered the young (spin-down age, $\tau$ = 63 kyr), energetic pulsar PSR J1105-6107 (J1105) which is located approximately 25' away from the remnant. It has a spin-down luminosity of 2.5$\times10^{36}$ erg s$^{-1}$ and overlaps the position of the $EGRET$ $\gamma$-ray source 3EG J1103-6106. It was also detected via periodicity searches in GeV $\gamma$-rays by the \textit{Fermi-LAT} satellite \citep{2013ApJS..208...17A}. It has a dispersion measure of 271 cm$^{-3}$ pc which implies a distance of $\sim$7 kpc using the standard Galactic electron density model \citep{2002astro.ph..7156C}. \citet{1997ApJ...485..820K} considered the scenario that this pulsar is associated with the remnant and determined from proper motion measurements that it would need to be travelling with a transverse velocity of $\sim$ 650 km s$^{-1}$ to have reached its current position, assuming a distance of 7 kpc to the pulsar and $\tau$ = 63 kyr. This is much larger than the average pulsar transverse velocity but much less than what has been suggested for other pulsar-SNR associations (e.g., \citealp{1993ApJ...415L.111C}), leading the authors to conclude that association is possible. Using the ASCA X-ray characteristics of MSH~11$-$6{\sl 1}A, \citet{2002ApJ...564..284S} concluded that MSH~11$-$6{\sl 1}A and J1105 are not associated under the assumption that the SNR evolved via thermal conduction or a cloudy ISM. These two models imply a transverse velocity of $\sim 4.5\times10^{3}$ km s$^{-1}$ and $\sim 5.3\times10^{3}$ km s$^{-1}$ respectively, which is much larger than the mean velocity ($\sim 310$ km s$^{-1}$) of young pulsars \citep{2005MNRAS.360..974H}. The two other pulsars, PSR J1103-6025 \citep{2003MNRAS.342.1299K} and PSR J1104-6103 \citep{1996AJ....111.2028K} are not associated with the remnant as they have characteristic ages much larger than 1 Myr, which is far greater than the expected lifetime of an SNR. Also nearby is the INTEGRAL source ICG J11014-6103, which is a neutron star travelling at a velocity exceeding 1000 km/s, which the \citet{2014A&A...562A.122P} associate with MSH~11$-$6{\sl 1}A.
  
Using 70 months of \textit{Fermi-LAT} data, we analyse the GeV $\gamma$-ray emission coincident with MSH~11$-$6{\sl 1}A and investigate the nature of this emission using broadband modelling. In addition, we analyse archival $Suzaku$ data and report on the spatial and spectral properties of the X-ray emission of this remnant. In Section 2 we describe how the \textit{Fermi-LAT} data are analysed and present the results of this analysis. In Section 3 we present our spatial and spectral analysis of the $Suzaku$ observation of MSH~11$-$6{\sl 1}A, while in Section 4 and 5 we discuss the implications of our results.

\section{OBSERVATIONS AND ANALYSIS OF MSH~11$-$6{\sl 1}A}

\begin{figure*}[htbp!]
\begin{center}
\includegraphics[width=0.49\textwidth]{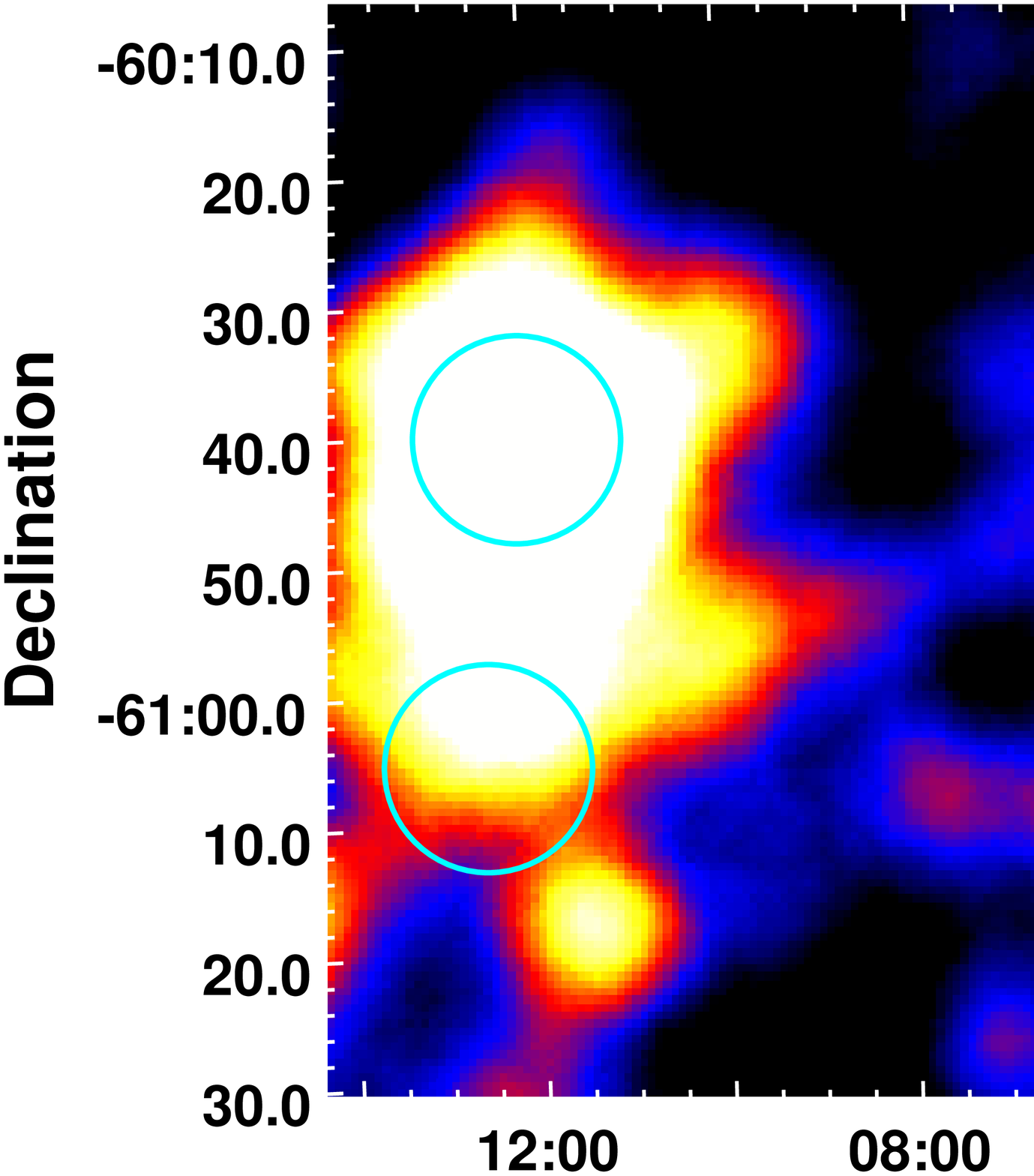}
\includegraphics[width=0.49\textwidth]{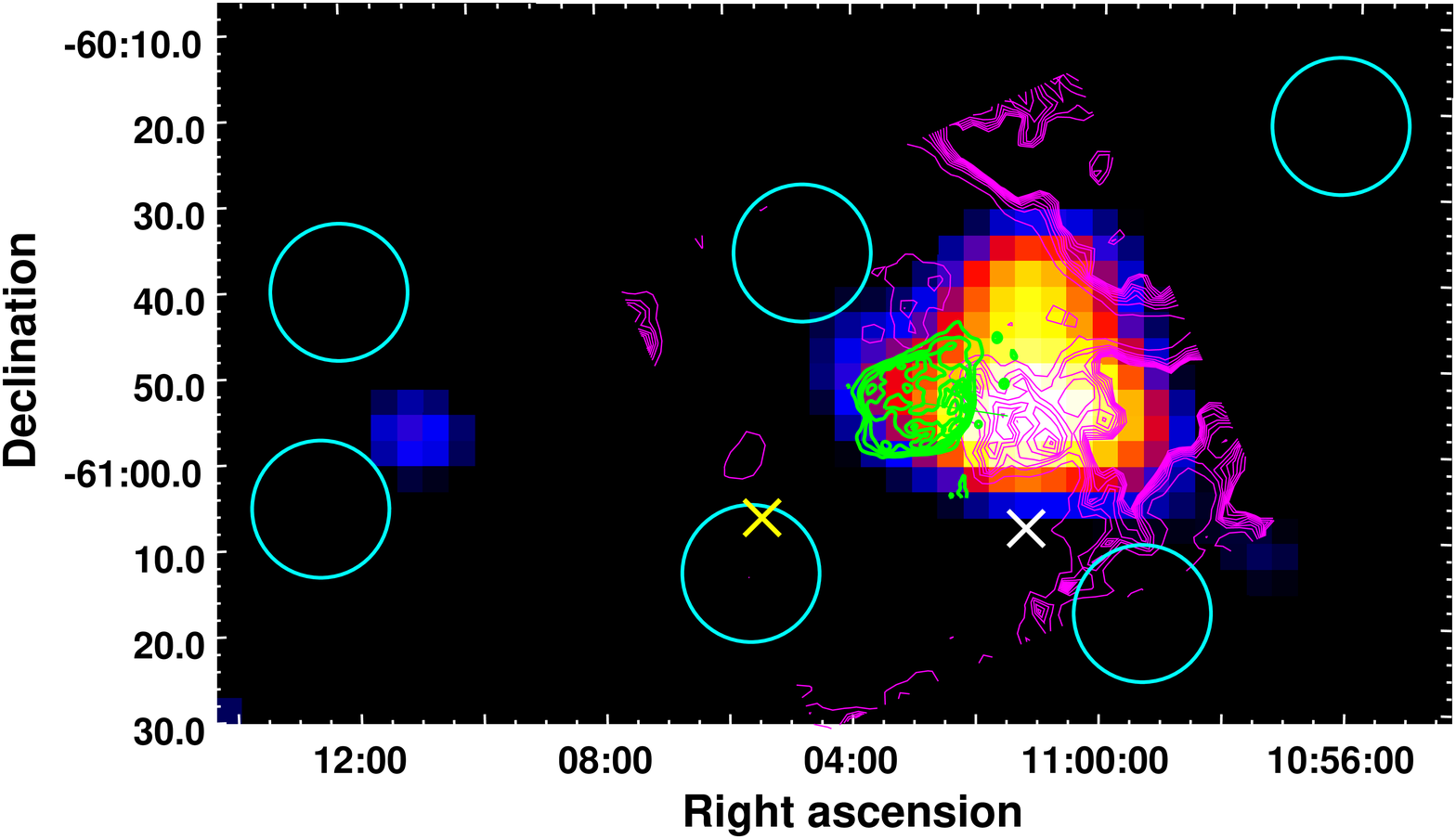}
\includegraphics[width=0.48\textwidth]{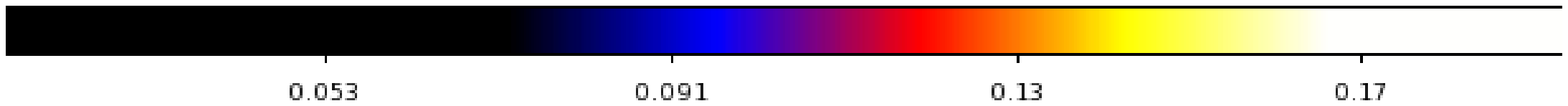}
\includegraphics[width=0.48\textwidth]{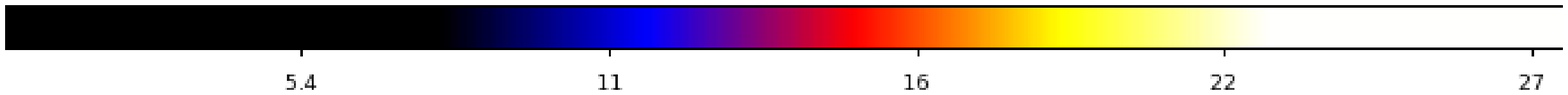}
\caption{\textit{Left:} $2.5^{\circ} \times 1.5^{\circ}$ \textit{Fermi-LAT} count map of the 2 - 200 GeV $\gamma$-ray emission surrounding MSH~11$-$6{\sl 1}A (units: counts degree$^{-2}$). The pixel binning is 0.01$^{\circ}$ and the maps are smoothed with a Gaussian of width 0.2$^{\circ}$.  The cyan circles correspond to the background \textit{Fermi-LAT} sources surrounding MSH~11$-$6{\sl 1}A. The brightest emission seen to the east of MSH~11$-$6{\sl 1}A corresponds to the $\gamma$-ray bright SNR MSH 11-62. The yellow $X$ corresponds to the radio position of PSR J1105-6107, while the white $X$ corresponds to the position of ICG J11014-6103 (the lighthouse nebula).  One can see that MSH~11$-$6{\sl 1}A is located in a complicated region of the $\gamma$-ray emitting sky. 843 MHz MOST radio continuum contours are overlaid in green. \textit{Right:} $2.5^{\circ} \times 1.5^{\circ}$ TS map of MSH~11$-$6{\sl 1}A. The magenta contours corresponds to the H$\textsc{i}$ emission of the molecular cloud associated with the remnant as detected by the Southern Galactic Plane Survey \citep{2005ApJS..158..178M}. The H$\textsc{i}$  emission contours range from 66 Kelvin at the edge of the molecular cloud interacting with MSH~11$-$6{\sl 1}A to 100 Kelvin at the center of the molecular cloud over seven linearly spaced contour levels. \label{countmap}} 
\end{center}
\end{figure*}
We analysed $\sim$70 months of reprocessed data collected by the \textit{Fermi-LAT} from 4th August 2008 to 16th June 2014. We selected data within a radius of 20$^{\circ}$ centered on MSH~11$-$6{\sl 1}A. We used the ``P7REP\_SOURCE\_V15'' instrument response function (IRF) which is based on the same in-flight event analysis and selection criteria that was used to generate the previous ``PASS7\_V6'' IRFs (details described in \citealt{2012ApJS..203....4A}). Due to the improved reconstruction of the calorimeter position as well as a 1\% per year correction for the degradation of the light yield of the calorimeter, the new IRFs significantly improve the point spread function (PSF) of the LAT for energy $>5$ GeV\footnote{More details found: \\
	 \textit{http://www.slac.stanford.edu/exp/glast/groups/canda/lat\_Performance.htm}}. 
The systematic uncertainties in the effective area of the \textit{Fermi-LAT} using ``P7REP\_SOURCE\_V15''  are 10\% below 100 MeV, decreasing logarithmically in energy to 5\% between 0.316 - 10 GeV and increasing logarithmically to 15\% at 1 TeV\footnote{\textit{http://fermi.gsfc.nasa.gov/ssc/data/analysis/LAT\_caveats.html}}. We selected events with a zenith angle less than 100$^{\circ}$ and that were detected when the rocking angle of the LAT was greater than 52$^{\circ}$ to decrease the effects of terrestrial albedo $\gamma$-rays. We analysed the $\gamma$-ray data in the direction of MSH~11$-$6{\sl 1}A using the Fermi Science Tools v9r33p0\footnote{\textit{http://fermi.gsfc.nasa.gov/ssc/data/analysis/software/}}
  
Due to the low count rates of $\gamma$-rays and the large PSF of the \textit{Fermi-LAT}, the standard maximum likelihood fitting technique, $gtlike$, was used to analyse the $\gamma$-ray emission of the remnant. Given a specific emission model, $gtlike$ determines the best-fit parameters of this model by maximising the joint probability of obtaining the observed data given an input model. $Gtlike$ accounts for background $\gamma$-ray emission by using diffuse Galactic and isotropic emission models described by the mapcube file \texttt{gll\_iem\_v05\_rev1.fits} and the isotropic spectral template \texttt{iso\_source\_v05\_rev1.txt}\footnote{The most up to date Galactic and isotropic emission models can be found \textit{http://fermi.gsfc.nasa.gov/ssc/data/access/lat/BackgroundModels.html}}. Gamma-ray emission from sources found in the Fermi-LAT second source catalogue are fixed to their position listed in the catalogue and their background contribution is calculated. The Galactic diffuse emission arises from the interaction of CRs with the interstellar medium and their subsequent decay into $\gamma$-rays, while the isotropic component arises from diffuse extragalactic $\gamma$-rays and residual charged particle emission. 
  
To improve the angular resolution of the data while analysing the spatial properties of the $\gamma$-ray emission of MSH~11$-$6{\sl 1}A, we selected $\gamma$-ray data converted in the \textit{front} section of the \textit{Fermi-LAT} with an energy range of 2 - 200 GeV. The improvement in spatial resolution in this energy range arises from the fact that the 1-$\sigma$ containment radius angle for $front$-selected photon events is $< 0.3^{\circ}$, while for lower energies it is much larger. To determine the detection significance, position and possible extent of the $\gamma$-ray emission coincident with MSH~11$-$6{\sl 1}A we produced test statistic (TS) maps using $gttsmap$ with an image resolution of 0.05$^{\circ}$. The TS is defined as $2log(L_{ps}/L_{null})$, where $L_{ps}$ is the likelihood of a point source being found at a given position on a spatial grid and $L_{null}$ is the likelihood of the model without the additional source.
  
To determine the spectral energy distribution (SED) of the $\gamma$-ray emission coincident with MSH~11$-$6{\sl 1}A we use events converted in the $front$ section of the LAT that have an energy of 0.2 - 204.8 GeV. This energy range is chosen to avoid the large uncertainties in the Galactic background model that arise below 0.2 GeV and to reduce the influence of the rapidly changing effective area of the LAT at low energies. We model the flux in each of 8 logarithmically spaced energy bins and estimate the best-fit parameters of the data using $gtlike$. We also include in the likelihood fit, background sources from the 24-month \textit{Fermi-LAT} second source catalogue \citep{2012ApJS..199...31N} that are found within 20$^{\circ}$ region centered on MSH~11$-$6{\sl 1}A. All evident background sources were identified in the \textit{Fermi-LAT} second source catalogue and the associated parameters from the catalogue were used. We left the normalization of the Galactic diffuse emission, isotropic component and the background point sources within 5 degrees of MSH~11$-$6{\sl 1}A free. For all other background point sources with a distance greater than 5 degrees from MSH~11$-$6{\sl 1}A their normalisations were frozen to that listed in the 2nd Fermi-LAT catalogue. In addition to the statistical uncertainties that were obtained from the likelihood analysis, systematic uncertainties associated with the Galactic diffuse emission were also calculated by artificially altering the normalisation of this background by $\pm$6\% from the best-fit value at each energy bin as outlined in \citet{2010ApJ...717..372C}.
  
In Figure \ref{countmap} left panel, we have generated a $2.5^{\circ} \times 1.5^{\circ}$ count map centered on MSH~11$-$6{\sl 1}A that is smoothed by a Gaussian with a width similar to the PSF for the events selected. MSH~11$-$6{\sl 1}A is located in a very complicated region of the sky. There are many \textit{Fermi-LAT} sources close to the remnant, with $\gamma$-ray bright SNR MSH 11-62 (2FGL J1112.1-6040 and 2FGL J1112.5-6105) located $\sim 1.2^{\circ}$, and PSR J1105-6107 (2FGL J1105.6-6114) located $\sim 0.36^{\circ}$, from the remnant. There are four other \textit{Fermi-LAT} sources in the immediate vicinity of MSH~11$-$6{\sl 1}A (2FGL J1104.7-6036, 2FGL J1105.6-6114, 2FGL J1059.3-6118c and 2FGL J1056.2-6021), but none of these are coincident with the MOST radio contours of the SNR. We define a source region at the position of MSH~11$-$6{\sl 1}A to estimate the flux from the SNR. Due to the close proximity of PSR J1105-6107 (J1105) and the low resolution of the \textit{Fermi-LAT} PSF ($\sim 1^{\circ}$ for front events at 68\% containment at $\sim$0.6 GeV), we can't rule out that this pulsar isn't contributing significantly to the observed $\gamma$-ray emission seen in Figure \ref{countmap}. Thus to analyse the spatial and spectral characteristics of the $\gamma$-ray emission of MSH~11$-$6{\sl 1}A, we have to remove the pulsar contribution. As MSH 11-62 is located $> 1^{\circ}$ away from MSH~11$-$6{\sl 1}A, it is unlikely that it is contributing significantly to the observed $\gamma$-ray emission.

\subsection{Removing the $\gamma$-ray contribution of PSR J1105-6107}

\begin{figure}[tbp!]
\begin{center}
\includegraphics[width=0.45\textwidth]{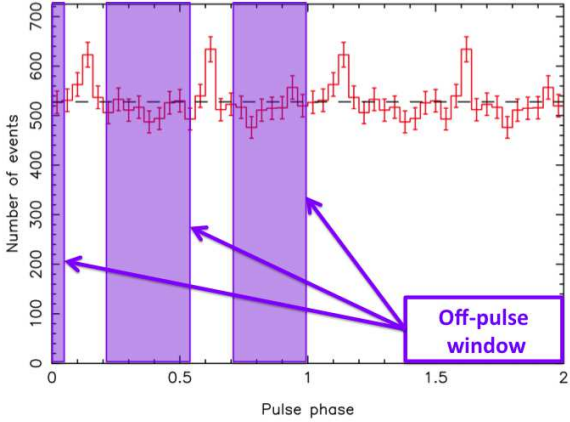}
\caption{The pulse-phase diagram of PSR J1105-6107 obtained using events in an energy range of 0.1 - 300 GeV coming from a 0.5$^{\circ}$ radius around the position of the pulsar. The off-pulse window used for this analysis is defined between 0.00-0.05, 0.22-0.55 and 0.70-1.00 of the pulse phase. Here two cycles are shown.\label{pulsephase}} 
\end{center}
\end{figure}

To avoid contamination from the pulsed emission of PSR J1105-6107 (J1105), we must perform our analysis in the off-pulse window of the pulsar light curve. The \textit{Fermi-LAT} collaboration made available with its second \textit{Fermi-LAT} catalogue of $\gamma$-ray pulsars \citep{2013MNRAS.429..688Y} the ephemerides of 117 $\gamma$-ray emitting pulsars that had been detected using three years of \textit{Fermi-LAT} data. Due to the continuous observations of the \textit{Fermi-LAT}, \citet{2013MNRAS.429..688Y} were able to directly determine regular times of arrival (TOAs) to produce a precise pulsar ephemeris.  We use the available ephemeris for J1105 which is valid from MJD 54200 (April 2007) to MJD 56397 (April 2013). As we are interested in analysing over $\sim$70 months of \textit{Fermi-LAT}, we need to check the validity of using the available J1105 ephemeris over the whole $\sim$70 months. Using the Parkes telescope, \citet{2013MNRAS.429..688Y} analysed J1105 searching for timing irregularities over a period of $\sim 16$ years from August 1994 to September 2010 (MJD 49589 - MJD 55461). J1105 experienced a glitch at MJD 50417 (November 1996),  MJD 51598 (February 2000), MJD 54711 (September 2008) and MJD 55288 (April 2010). The \textit{Fermi-LAT} ephemeris of J1105 covers the glitches experienced by the pulsar since the beginning of the \textit{Fermi-LAT} mission. As of writing, there have been no other reports in the literature that J1105 has undergone glitches since April 2010. We also tested whether the pulsar is noisy in $\gamma$-rays, as this can also indicate irregularities in the pulsar rotation that have not been incorporated in the ephemeris. We produced a pulse phase vs. events plot during the period of validity of the ephemeris (MJD 54200 to MJD 56397) and for the period between the end time of the ephemeris and the end date of our data (MJD 56397 to MJD 56824) to see if the pulse peaks that we observe in Figure \ref{pulsephase} disappear due to noise of the pulsar. We find that during both time periods we obtain the same pulse phase profile. As no glitches have been detected as of writing and the pulsar has not been noisy beyond MJD 56397, we assume that this ephemeris is valid for our whole data set.
  
The $\gamma$-ray photons were phase-folded using this ephemeris and the pulse phases were assigned to the \textit{Fermi-LAT} data using the \textit{Fermi-LAT} TEMPO2 plugin provided by the \textit{Fermi-LAT} collaboration\footnote{\textit{http://fermi.gsfc.nasa.gov/ssc/data/analysis/user/Fermi\_plug\_doc.pdf}}. This plugin calculates the rotational phase of the pulsar for each photon arrival time in the \textit{Fermi-LAT} data using the barycentric dates of each event. Using $ftselect$, we remove the pulse and use only the $\gamma$-ray photons in the off-pulse window (defined by the 0.00-0.05, 0.22-0.55 and 0.70-1.00 pulse phase intervals) to perform our spectral and morphological analysis. In Figure \ref{pulsephase} we have plotted the pulse-phase diagram of J1105 obtained in the 0.10-300 GeV energy range using 0.5$^{\circ}$ radius around the position of J1105.

\subsection{TS Map}
In Figure \ref{countmap} right panel we present a TS map of MSH~11$-$6{\sl 1}A using the off-pulse $\gamma$-ray data. This was calculated using \textit{gttsmap} over an energy of 0.2 - 2.0 GeV and using \textit{front} events only. In addition to the diffuse Galactic background components, we include in the background model the \textit{Fermi-LAT} sources associated with MSH 11-62 (2FGL J1112.1-6040 and 2FGL J1112.5-6105) and the four sources in the immediate vicinity of the remnant (2FGL J1104.7-6036, 2FGL J1105.6-6114, 2FGL J1059.3-6118c and J1056.2-6021). The TS map suggests that there is significant $\gamma$-ray emission coincident with MSH~11$-$6{\sl 1}A and the MC associated with the remnant, as highlighted by the magenta contours. The peak of the $\gamma$-ray emission is found at a best fit position of $(\alpha_{J2000},\delta_{J2000}) = (11^{h}01^{m}{29}^{s}, -60^{\circ}55’29’’)$, placing it outside the SNR boundary, but consistent with being located along or inside the western limb given the angular resolution of the \textit{Fermi-LAT}. The emission is detected with a significance of $\sim 5\sigma$, with the $\gamma$-ray emission at the center of the remnant producing a significance of $\sim4\sigma$. One can see in Figure \ref{countmap} right panel that the contribution of MSH 11-62 and the other \textit{Fermi-LAT} sources surrounding the remnant have been modelled out, while the contribution from the pulsar (J1105) has been gated out successfully.

\subsection{$\gamma$-ray spectrum}
\begin{figure}[t]
\begin{center}
\includegraphics[width=0.45\textwidth]{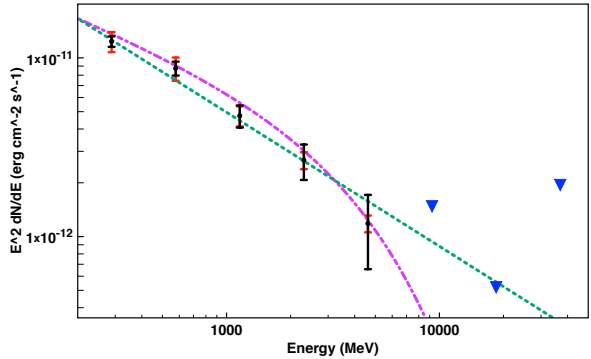}
\caption{The $Fermi-LAT$ $\gamma$-ray spectrum of MSH~11$-$6{\sl 1}A. Statistical errors are shown as black error bars, systematic errors are plotted as red error bars, while the upper limits are plotted as blue triangles. The simple power law model with $\Gamma = 2.75^{+0.07}_{-0.06}$ and exponential cut-off power law model with $\Gamma = 2.49^{+0.17}_{-0.19}$ and $E_{cut} =4.20^{+1.91}_{-0.66}$ GeV are shown as the green dotted and magenta dot-dashed line respectively. \label{gammaspectrum}} 
\end{center}
\end{figure}

The $\gamma$-ray spectrum of MSH~11$-$6{\sl 1}A is shown in Figure \ref{gammaspectrum}, with the statistical errors plotted in black and systematic errors plotted in red. For energies above 6.40 GeV, only flux upper limits have been determined and are plotted as blue triangles. Additionally, the best fit power law and exponential cut-off power law models are plotted as the green dotted line and purple dot-dashed line respectively. The $\gamma$-ray spectrum can be fit using a simple power law with a spectral index of $2.75^{+0.07}_{-0.06}$, giving a reduced $\chi^{2} \sim 1$. An exponential cut-off power law with $E_{cut} =4.20^{+1.91}_{-0.66}$ GeV and spectral index of $2.49^{+0.17}_{-0.19}$ can fit the spectrum equally well giving a reduced $\chi^{2} \sim 0.8$ for the fit. For an energy range of 0.1-100 GeV, the integrated flux is $(4.2^{+0.34}_{-0.98})\times10^{-11}$ erg cm$^{-2}$s$^{-1}$, assuming the power law fit. Using a distance of 7 kpc, the luminosity of this $\gamma$-ray source in this energy range is $(2.5^{+0.17}_{-0.63})\times10^{35}$ erg s$^{-1}$.

\section{$Suzaku$ OBSERVATIONS OF MSH~11$-$6{\sl 1}A}

MSH~11$-$6{\sl 1}A was observed with $Suzaku$ using X-ray imaging spectrometers (XIS) \citep{2007PASJ...59S..23K} on 2011 July 25th for $\sim 111$ks (ObsID 506061010).  For this observation only XIS0\footnote{It is also important to note that a fraction of XIS0 has not been functional since 2009 June 23rd due to the damage caused by a micro meteorite. For more information see: \textit{http://www.astro.isas.jaxa.jp/suzaku/doc/suzakumemo/suzakumemo-2010-01.pdf}.}, XIS1 and XIS3 observations are available as XIS2 has not been functional since November 2006\footnote{\textit{http://www.astro.isas.jaxa.jp/suzaku/doc/suzakumemo/suzakumemo-2007-08.pdf}}. Recently, \citet{2014arXiv1411.6809K} presented their analysis of this $Suzaku$ observation. They found recombining plasma in the center and in the northwest of the remnant which has enhanced abundances and a temperature of $\sim$0.5 keV, while everywhere else the X-ray emission arises from an ionising ISM component with a temperature of $\sim$0.6 keV. In section 5.2.1, we estimate the density of the $\gamma$-ray emitting material based on our \textit{Fermi-LAT} spectrum (Figure \ref{gammaspectrum}). To test whether this inferred density agrees with other observations, we have re-analysed the $Suzaku$ data in order estimate the density of the surrounding environment.

For our analysis we used the standard tools of $HEASOFT$ version 6.16.  We reprocessed the unfiltered public data using $aepipeline$ (version 1.1.0) and use the current calibration database (CALDB) available as of 2014 July 1st (version 20140701). Following the standard screening criteria\footnote{\textit{http://heasarc.nasa.gov/docs/suzaku/processing/criteria xis.html}}, we filtered hot and flickering pixels, time intervals corresponding to $Suzaku$ passing the South Atlantic Anomaly and night-earth and day-earth elevation angles less than 5$^{\circ}$ and 20$^{\circ}$, respectively. We utilised events that had a grade of 0, 2, 3, 4 and 6 only. The total exposure of our observation is 111 ks for each of the XIS detectors. We extracted the spectra and images of the remnant from the $5\times5$ and $3\times3$ editing mode event files using XSELECT version 2.4. For the spectral analysis we generated the redistribution matrix file (RMF) and ancillary response files (ARF) using $XISRMFGEN$ and $XISARFGEN$ respectively. To analyse the spectral data we used the X-ray spectral fitting package (XSPEC) version 12.8.2q with AtomDB 3.0.1\footnote{AtomDB 3.0.1 can be downloaded here: \textit{http://www.atomdb.org/download.php}} \citep{2001ApJ...556L..91S, 2012ApJ...756..128F}. 

\subsection{Spectral analysis of the individual annulus and rectangular regions}
\begin{figure*}[tbp!]
\begin{center}
\includegraphics[width=0.49\textwidth]{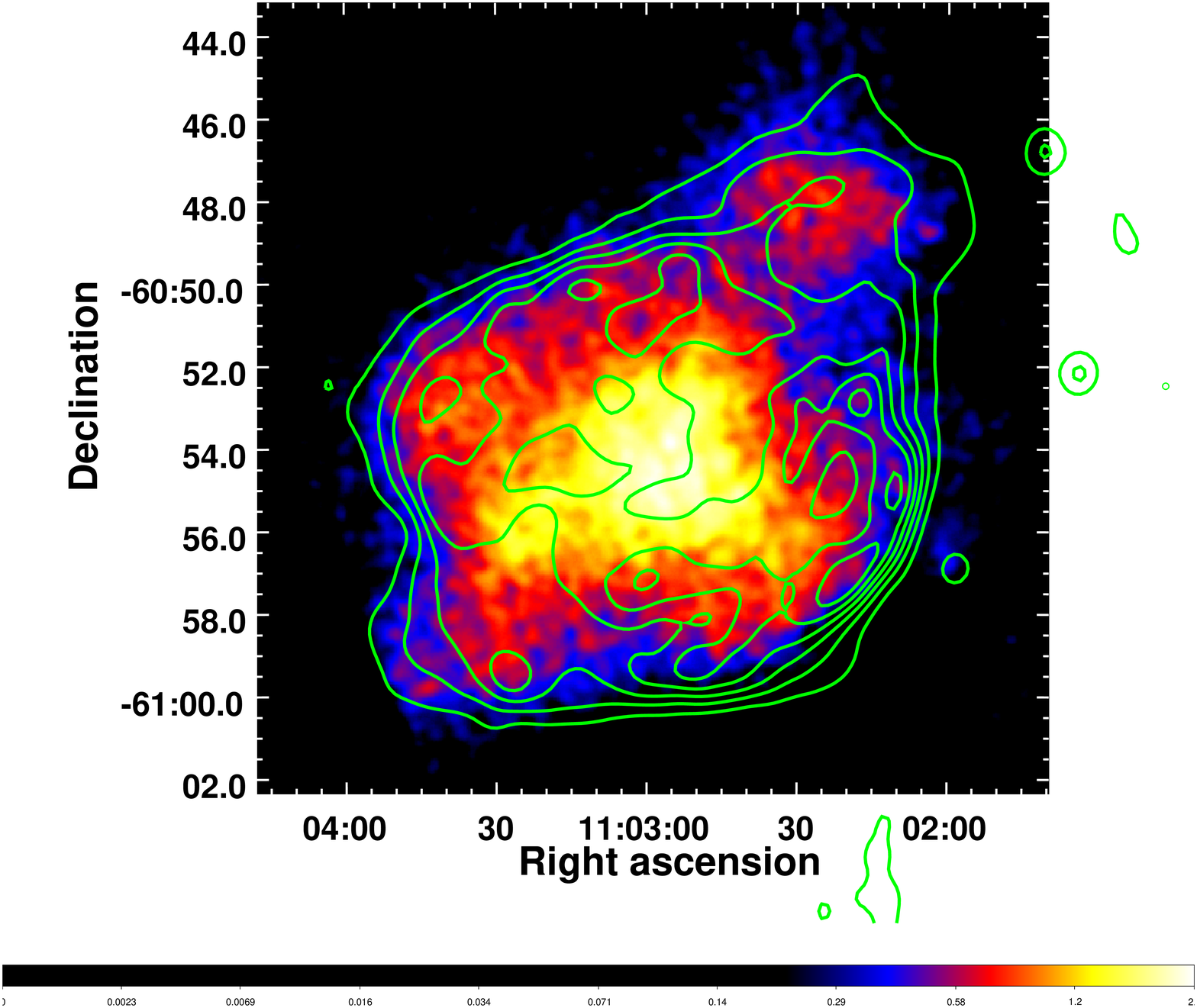}
\includegraphics[width=0.49\textwidth]{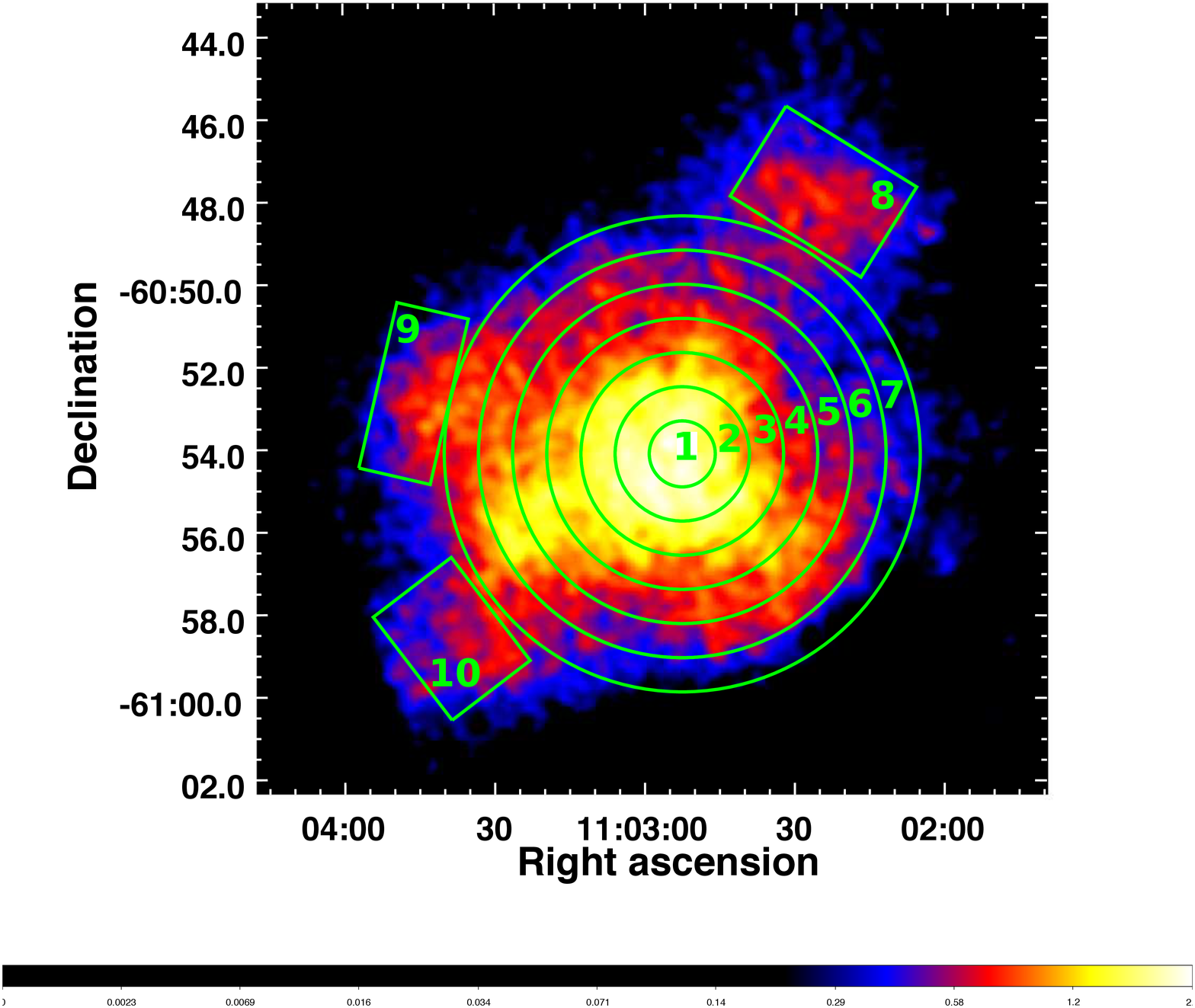}
\caption{$Suzaku$ XIS0 image of MSH~11$-$6{\sl 1}A in the 0.5-7.0 KeV energy band.  The image has been smoothed with a Gaussian function of width 0.1' and has a logarithmic scaling applied to it. $Left:$ The green contours correspond to the 843 MHz MOST radio continuum emission used in Figure \ref{countmap}. $Right:$ Overlaid in green are the regions we use for spectral extraction, described in Section 3.1.  \label{suzaku}} 
\end{center}
\end{figure*}
\begin{table}[tbp!]
\begin{center}
\caption{Best fit parameters for our background model. \label{bkg}}
    \begin{tabular}{ccc}
    \hline
Component & Parameter & Value \\
  \hline\hline
\textit{Cosmic X-ray background} &  N$_{H} (\times 10^{22}$ cm$^{-2}$)	&$	1.47	_{	-0.83	}^{+	1.07	}$\\
 & $\Gamma$	&1.40 (frozen)\\
  \hline
\textit{Galactic ridge emission}  &N$_{H} (\times 10^{22}$ cm$^{-2}$)	&$	0.61_{	-0.06	}^{+	0.08	}$\\
&kT (keV)	&$	0.22\pm0.02$\\
 &Abundances	&	0.20 (frozen)\\
 \hline
\textit{Galactic halo} &N$_{H} (\times 10^{22}$ cm$^{-2}$)	&$	1.64_{	-0.10	}^{+	0.13	}$\\
&kT (keV)	&$	0.58\pm0.02$\\
&Ne &$	2.71	_{	-0.40	}^{+	0.60	}$\\
&Mg	&$	2.31_{	-0.18	}^{+	0.23	}$\\
&Si	&$	3.48	_{	-0.32	}^{+	0.34	}$\\
\hline
&Reduced $\chi^2$ (dof)	&	1.80 (1602)	\\			
\hline
    \end{tabular}
        \end{center}
        \begin{tablenotes}
      \small
      \item   Note: All uncertainties correspond to 1$\sigma$ errors.
    \end{tablenotes}
\end{table}
\begin{table*}[htbp!]
\caption{ Spectral fits for all 10 individual regions.\label{fits}}

 \begin{center}
    \begin{tabular}{ccccccccccccccccccccccc}
    \hline
Region	&	$N_{H}$ ($10^{22}$ cm$^{-2}$)					&	kT	(keV)				&	$kT_{init}$ (keV)	&	Ne					&	Mg					&	Si					&	S					&	Fe					&	$\tau$  ($10^{12} $ s cm$^{-3}$)					&	Reduced $\chi^2$	\\
\hline
1	& $	1.76	_{	-0.15	}^{+	0.12	} $ & $	0.43	_{	-0.02	}^{+	0.03	} $ & $	5	 $ & $	-					$ & $	3.23	_{	-0.46	}^{+	0.52	} $ & $	6.32	_{	-0.92	}^{+	1.07	} $ & $	1.92	_{	-0.19	}^{+	0.24	} $ & $	-					$ & $	1.59	_{	-0.14	}^{+	0.19	} $ &	0.79	\\
2	& $	1.20	_{	-0.11	}^{+	0.12	} $ & $	0.42	_{	-0.02	}^{+	0.02	} $ & $	5	 $ & $	0.23	_{	-0.09	}^{+	0.12	} $ & $	1.82	_{	-0.20	}^{+	0.26	} $ & $	4.96	_{	-0.34	}^{+	0.38	} $ & $	2.51	_{	-0.34	}^{+	0.38	} $ & $	0.11	_{	-0.06	}^{+	0.08	} $ & $	1.29	_{	-0.06	}^{+	0.08	} $ &	0.94	\\
3	& $	1.32	_{	-0.13	}^{+	0.14	} $ & $	0.40	_{	-0.02	}^{+	0.03	} $ & $	5	 $ & $	0.35	_{	-0.12	}^{+	0.17	} $ & $	2.04	_{	-0.26	}^{+	0.33	} $ & $	4.65	_{	-0.42	}^{+	0.50	} $ & $	2.62	_{	-0.36	}^{+	0.43	} $ & $	0.14	_{	-0.08	}^{+	0.11	} $ & $	1.26	_{	-0.06	}^{+	0.06	} $ &	1.02	\\
4	& $	1.81	_{	-0.09	}^{+	0.24	} $ & $	0.34	_{	-0.09	}^{+	0.02	} $ & $	5	 $ & $	-					$ & $	2.93	_{	-0.28	}^{+	0.33	} $ & $	5.69	_{	-0.51	}^{+	0.73	} $ & $	3.72	_{	-0.71	}^{+	0.41	} $ & $	-					$ & $	1.34	_{	-0.07	}^{+	0.04	} $ &	1.08	\\
5	& $	1.87	_{	-0.17	}^{+	0.10	} $ & $	0.27	_{	-0.01	}^{+	0.06	} $ & $	5	 $ & $	-					$ & $	2.49	_{	-0.28	}^{+	0.33	} $ & $	5.21	_{	-0.75	}^{+	0.74	} $ & $	3.34	_{	-1.14	}^{+	1.52	} $ & $	-					$ & $	1.28	_{	-0.11	}^{+	0.09	} $ &	1.06	\\
6	& $	1.66	_{	-0.08	}^{+	0.04	} $ & $	0.30	_{	-0.02	}^{+	0.03	} $ & $	5	 $ & $	-					$ & $	2.67	_{	-0.20	}^{+	0.51	} $ & $	3.69	_{	-0.41	}^{+	0.67	} $ & $	2.12	_{	-0.44	}^{+	0.46	} $ & $	-					$ & $	1.00	_{	-0.08	}^{+	0.09	} $ &	1.10	\\
7	& $	1.67	_{	-0.16	}^{+	0.21	} $ & $	0.36	_{	-0.03	}^{+	0.05	} $ & $	5	 $ & $	-					$ & $	4.11	_{	-0.77	}^{+	1.07	} $ & $	7.05	_{	-1.30	}^{+	1.92	} $ & $	4.27	_{	-1.11	}^{+	1.42	} $ & $	-					$ & $	1.09	_{	-0.07	}^{+	0.08	} $ &	1.14	\\
8	& $	2.27	_{	-0.21	}^{+	0.23	} $ & $	0.42	_{	-0.04	}^{+	0.04	} $ & $	5	 $ & $	-					$ & $	2.59	_{	-0.46	}^{+	0.58	} $ & $	4.87	_{	-0.86	}^{+	1.11	} $ & $	2.91	_{	-0.66	}^{+	0.83	} $ & $	-					$ & $	0.90	_{	-0.06	}^{+	0.05	} $ &	0.92	\\
9	& $	2.00	_{	-0.30	}^{+	0.45	} $ & $	0.80	_{	-0.27	}^{+	0.22	} $ & $	0.0808	 $ & $	-					$ & $	3.63	_{	-1.02	}^{+	2.77	} $ & $	6.91	_{	-1.96	}^{+	5.72	} $ & $	3.06	_{	-1.17	}^{+	5.20	} $ & $	-					$ & $	0.22	_{	-0.11	}^{+	0.22	} $ &	0.84	\\
10	& $	2.46	_{	-0.37	}^{+	0.38	} $ & $	0.73	_{	-0.14	}^{+	0.24	} $ & $	0.0808	 $ & $	-					$ & $	4.69	_{	-1.03	}^{+	1.67	} $ & $	6.67	_{	-1.43	}^{+	2.29	} $ & $	2.86	_{	-0.90	}^{+	1.43	} $ & $	-					$ & $	0.19	_{	-0.11	}^{+	0.39	} $ &	0.94	\\
\hline
\hline
    \end{tabular}
        \begin{tablenotes}
      \small
      \item    Note:  All uncertainties correspond to the 90\% confidence level.
    \end{tablenotes}

    \end{center}

\end{table*}

We extracted spectra from a central circular region defined by region 1 in Figure 4 (right panel) and 6 annular regions of width 0.82$^\prime$ to cover the central X-ray emission of the remnant (regions 2 - 7 in Figure 4, right panel). The radial size of these regions was chosen to be the same size as the angular resolution of $Suzaku$	($\sim 0.8^\prime$).  These regions were chosen to fully enclose the bright central X-ray emission of the remnant, which is quite symmetric in nature.  We also extracted spectra from three rectangular regions (regions 8 - 10 in Figure 4, right panel) that are not covered by the annulus regions, to enclose protrusions in the northwest, southeast, and east. Annular regions were chosen in order to characterise radial variations in the brightness, temperature, ionisation state and elemental abundances of the remnant, all of which are important for understanding the nature of the mixed morphology. Our choice of regions differs from those chosen by \citet{2014arXiv1411.6809K}, who also analysed the Suzaku observation of MSH~11$-$6{\sl 1}A. They extracted spectra from five regions that do not enclose the full X-ray emission from the remnant -- a central region that corresponds to our regions 1, 2, and 3; a northwest region that encompasses our region 8 and a northwestern portion of region 7; and NE, SE, and SW regions that cover sectors of our annular regions and also encompass our region 9 in the east. All spectra were grouped by 20 counts using the FTOOLS command $grppha$ and all available XIS detectors were used.
  
To estimate the background, we extract data from the full field of view of the XIS of our observation, excluding the calibration regions and the emission from the remnant. The background spectrum consists of two major components, the non X-ray background and the astrophysical background which is made up of the cosmic X-ray background, the Galactic ridge X-ray emission and the Galactic halo. We use $xisnxbgen$ \citep{2008PASJ...60S..11T}  to generate a model for the NXB which we then subtract from our background spectrum. Similarly, we subtract a model for the NXB from our spectra obtained from the regions shown in Figure \ref{suzaku} right panel. Similar to \citet{2014arXiv1411.6809K}, we model this NXB subtracted spectrum. We fix the cosmic X-ray background power law component to that of \citet{2002PASJ...54..327K}, and use a single apec model with a temperature and surface brightness ($\sim 1.0\times10^{-12}$ erg cm$^{-2}$ s$^{-1}$ deg$^{-2}$ for 0.5-2.0 keV) similar to that obtained by \citet{2013ApJ...773...92H} to define our Galactic halo component. We use a single low temperature apec model with subsolar abundances frozen to that of \citet{1997ApJ...491..638K} to define the Galactic ridge emission. We also use the Wilms et al. abundance table \citep{2000ApJ...542..914W}. Our best-fit parameters for our background and their uncertainties are given in Table \ref{bkg}.
  
To model the X-ray emission from the remnant we used a non-equilibrium ionisation (NEI) collisional plasma model, VVRNEI, which is characterised by a final ($kT$) and initial electron temperature ($kT_{init}$), elemental abundances and a single ionisation timescale ($\tau = n_{e}t$). This allows one to model a plasma that is ionising up to collisional equilibrium from a very low initial temperature $kT_{init}$, mimicking the standard NEI/VNEI model that is commonly used in the literature. Additionally, the RNEI/VVRNEI model can reproduce a recombining (overionised) plasma where one assumes that the plasma starts in collisional equilibrium with an initial temperature $kT_{init}$ that suddenly drops to its final temperature $kT$. For our analysis, the column density, ionisation timescale, normalisation, and final temperature were left as free parameters. Due to the strong emission lines from Mg, Si, and S the abundances of these elements were also left free. Additionally we also let Ne and Fe be free parameters for regions 2 - 3, as we found that varying these significantly improved the fit. All other elemental abundances were frozen to the solar values reported by \citet{2000ApJ...542..914W}. The foreground absorbing column density $N_{H}$ was modelled using TBABS \citep{2000ApJ...542..914W}. Figure \ref{spectrum} shows an example of the X-ray spectrum of MSH~11$-$6{\sl 1}A as extracted from region 3. The spectra derived for each region shown in Figure \ref{suzaku} right panel all have similar features to the spectrum shown in Figure \ref{spectrum}.
  
We found that for regions 1 - 8 the fit favoured an initial temperature larger than the final temperature, while regions 9 and 10 favoured an initial temperature smaller than the final temperature. When left free, these initial temperatures would hit the upper (or lower) limits of this parameter and the associated abundances we obtained for our fits were unrealistically high (abundances of [Mg], [Si] > 10 relative to \citealt{2000ApJ...542..914W}). The ionisation timescale for all regions was $\tau \sim 10^{12}{\rm\ s\ cm}^{-3}$. 

\begin{figure}[tbp!]
\begin{center}
\includegraphics[width=0.49\textwidth]{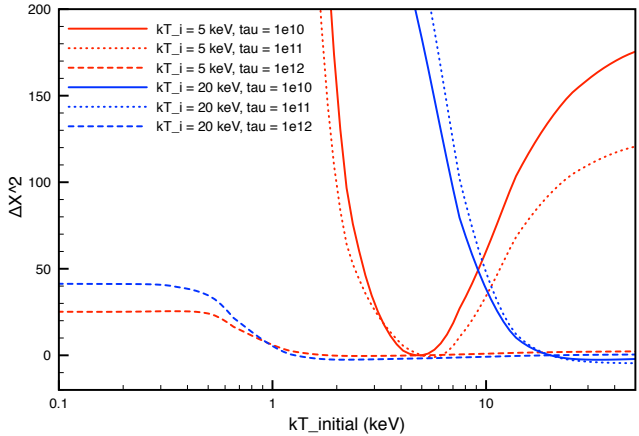}
\caption{The sensitivity of our fits to changes in $kT_{init}$ based on $\Delta\chi^{2}$. Here the red plots correspond to $kT_{init}$ = 5 keV, and the blue corresponds to $kT_{init}$ = 20 keV. The solid, dotted and dot-dashed lines correspond to $\tau$ = 10$^{10}$  s cm$^{-3}$, 10$^{11}$  s cm$^{-3}$ and 10$^{12}$  s cm$^{-3}$ respectively. We assume $kT$ = 0.5 keV and $n_{H} = 10^{22}$ cm$^{-2}$ for all spectra. From this plot we derive an upper limit for $kT_{init}$ of 2 - 5 keV corresponding to a shock velocity of 1300 - 2100 km/s, for which we then use in our fits.}\label{sensitivity}
\end{center}
\end{figure}

To investigate the sensitivity of our fits to values of $kT_{init}$, we simulated spectra with similar counting statistics to those from our regions of investigation, using $kT = 0.5$~keV, $N_H = 10^{22}{\rm\ cm}^{-2}$, and $kT_{init} = 5$~keV or 20~keV. We considered cases for $\log \tau =$~10, 11, and 12. We fit each spectrum to a TBABS*VVRNEI model and then investigated the effect of freezing $kT_{init}$ values over a range from 0.1 to 100~keV. The results are illustrated in Figure \ref{sensitivity} where we plot $\Delta \chi^2$ vs. $kT_{init}$ for spectra with actual $kT_{init}$ values of 5~keV (red) and 20~keV (blue). Here the solid, dotted, and dashed lines correspond, respectively, to $\log \tau$ =~10, 11, and 12. For low ionisation timescales the resulting fits are quite sensitive to the $kT_{init}$ value, while for timescales similar to that found in MSH~11$-$6{\sl 1}A\ ($\tau \approx 10^{12}{\rm\ s\ cm}^{-3}$), the fits are insensitive to $kt_{init}$ (i.e. the $\Delta \chi^2$ vs. $kT_{init}$ plot plateaus) for temperatures greater than 2-5 keV. This is because above $\sim 2 - 5$~keV, it is only the emission from Fe and Ni that is significantly impacted by higher $kT_{init}$ values, because all other abundant ions are fully stripped at these temperatures. Our observations do not have enough counts around the Fe and Ni lines to provide sensitivity to such an effect (though longer observations, particularly with higher resolution, would provide such sensitivity; see below).

Based on where $\Delta \chi^2$ vs. $kT_{init}$ plot plateaus for $\tau \approx 10^{12}{\rm\ s\ cm}^{-3}$, we fix value of $kT_{init}$ at 5~keV for all regions other than 9 and 10. For the latter regions, whose fits indicate $kT_{init} < kT$), we fix the initial temperature at the minimum available value for the VVRNEI model (80.8~eV). In Table \ref{fits}, we list the best fit parameters for each of our spectra.  All uncertainties correspond to the 90\% confidence level.

A recombining plasma that has an initial temperature of 2 - 5 keV implies that the shock must have had a velocity of $\sim$ 1300 - 2100 km s$^{-1}$, assuming electron-ion equilibrium. This high initial velocity suggests that the recombining plasma	could be the result of the SNR shock front initially expanding into a dense CSM, reaching a high ionisation state corresponding to the	high initial expansion velocity (and, thus, shock temperature). Such a scenario might result from expansion into an $r^{-2}$ density	profile characteristic of a stellar wind, with subsequent expansion	into the lower density regions resulting in rapid cooling, leaving	an overionised plasma \citep{1989MNRAS.236..885I, 2012ApJ...750L..13M}. 

\begin{figure}[tbp!]
\begin{center}
\includegraphics[width=0.49\textwidth]{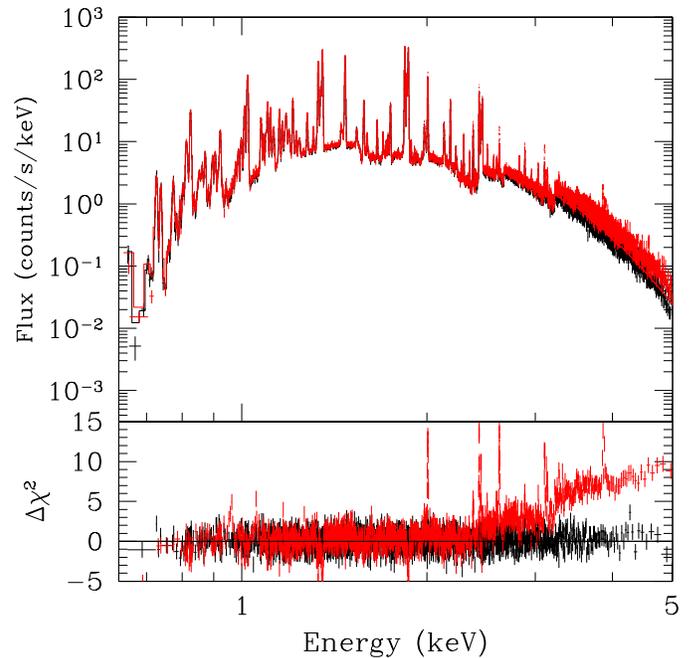}
\caption{A 20 ks background subtracted spectrum simulated for \textit{ASTRO-H} for a recombining plasma that has an initial temperature of 2 keV (black) and one that has an initial temperature of 50 keV (red). The model was based on that obtained for region 8 but with the initial temperature set to 2 keV and 50 keV. With this observation one can easily differentiate between a recombining plasma that has two different initial temperatures.}\label{astroh}
\end{center}
\end{figure}

\begin{figure*}[tbp!]
\begin{center}
\includegraphics[width=0.8\textwidth]{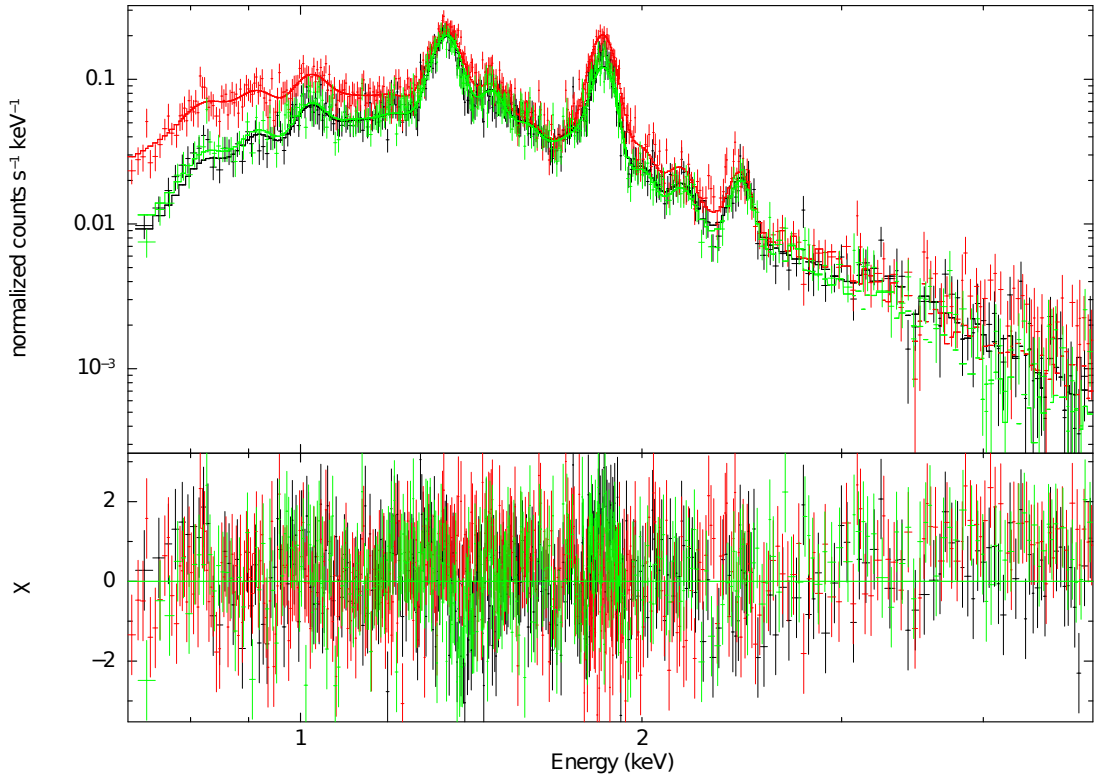}
\caption{The $Suzaku$ XIS0, XIS1 and XIS3 spectrum extracted from region 3 in Figure \ref{suzaku}. These spectra were fitted simultaneously using our background model described in Table \ref{bkg} and an absorbed VVRNEI model as described by the parameters listed in Table \ref{fits}. The spectra are overlaid with the best fit model with their $chi^2$ residuals. The spectra derived for each region shown in Figure \ref{suzaku} right panel all have similar features as the spectrum shown here.\label{spectrum}} 
\end{center}
\end{figure*}

Even though our current data are unable to differentiate between a plasma that has an initial temperature of 2 - 5 keV or one that has initial temperature greater than this, with the launch of \textit{ASTRO-H} we will be able to differentiate between these two cases. In Figure \ref{astroh} we have plotted a simulated background-subtracted spectrum that we would obtain in a 20~ks observation using the calorimeter on \textit{ASTRO-H} for a plasma that has an initial temperature of 2 keV and one that has an initial temperature of 50 keV, assuming our fit parameters for region 8. The background spectrum for \textit{ASTRO-H} was obtained from SIMX simulations\footnote{\textit{https://hea-www.harvard.edu/simx/}}. Here the black spectrum corresponds to a plasma that has an initial temperature of 2 keV (or an initial velocity of $\sim$1300 km s$^{-1}$), while the red spectrum corresponds to a plasma that has an initial temperature of 50 keV (or an initial velocity of $\sim$7000 km/s, typical of the high initial expansion speed of an SNR). One can see that with a 20 ks \textit{ASTRO-H} observation, we could easily differentiate between two plasmas that have different initial temperatures.

Our spectra from MSH~11$-$6{\sl 1}A are best described by a recombining plasma model with the exception of emission from regions 9 and 10, in the eastern and southeastern outskirts, where an ionising plasma is observed. Our results for the central regions (1 - 3) agree well with those of \citet{2014arXiv1411.6809K}, who also find a recombining plasma for their southwestern region, in agreement with our results. In contrast, for their southeastern and northeastern regions, \citet{2014arXiv1411.6809K} obtain best fits for an ionising plasma. Given that these regions combine emission from the eastern and southeastern protrusions (our regions 9 and 10), for which we also observe an ionising plasma, with emission from the outer symmetric portion of the SNR, where we observe a recombining plasma, the combination of two components may explain the partial discrepancy. Conversely, \citet{2014arXiv1411.6809K} report an ionising plasma for their southwestern region, which covers regions for which we obtain  a recombining plasma. This may result from our annular regions averaging over multiple components.
	
All regions have an ionisation timescale of $\sim 1.0\times10^{12}$ cm$^{-3}$ s, indicating that the X-ray emitting plasma across the whole remnant is close to ionisation equilibrium \citep{2010ApJ...718..583S}. The temperature of the recombining components ranges from 0.27 keV to 0.43 keV near the bright central X-ray emission. Interestingly, the regions that are best described by an ionising plasma have the highest temperatures out of all the regions that we analysed, with temperatures of 0.80 keV and 0.73 keV respectively. On average, our derived temperatures are lower than that derived by \citet{2014arXiv1411.6809K}, \citet{2012A&A...546A..91G} in their $XMM-Newton$ analysis, and \citet{2002ApJ...564..284S} in their $ASCA$ analysis.

We find in all regions strong emission lines coming from Mg, Si and S and all regions require super-solar abundances of these elements. The enhancement of elemental abundances is observed in many MM SNRs and indicates that some of the X-ray emission we are observing arises from ejecta that have been dispersed throughout the remnant and been mixed with the swept up shocked material. Similar to \citet{2014arXiv1411.6809K}, and \citet{2012A&A...546A..91G} we also find an underabundance of Ne and Fe in regions 2 - 3. Unlike, \citet{2014arXiv1411.6809K} we do not find evidence for overabundance of Ar or the underabundance of O suggested by \citet{2012A&A...546A..91G}. When we freed these parameters we found that they do not significantly improve our fit, thus we kept them at solar. Our estimated abundances are slightly higher than that derived by \citet{2014arXiv1411.6809K}, \citet{2012A&A...546A..91G} and \citet{2002ApJ...564..284S}. This discrepancy arises from the fact that in our analysis we use the abundance table by \citet{2000ApJ...542..914W} and the newly updated ATOMDB 3.0.1, while \citet{2014arXiv1411.6809K} and \citet{2012A&A...546A..91G} use the table derived by \citet{1989GeCoA..53..197A} and ATOMDB 3.0 and ATOMDB 2.0.2 respectively.

Our derived column density towards MSH~11$-$6{\sl 1}A ranges between $(1.20_{-0.11}^{+0.12}-2.46_{-0.37}^{+0.38})\times 10^{22}$ cm$^{-2}$. The column density is highest in regions 8 - 10 which directly interacts with the surrounding environment. Our estimates for $N_{H}$ are higher than the column density derived by \citet{2002ApJ...564..284S}, \citet{2012A&A...546A..91G} and \citet{2014arXiv1411.6809K}. This discrepancy arises from the fact that we use a different abundance table and a different absorption model.

\subsection{Deriving the density of the X-ray emitting gas}

\begin{table}[t]
\caption{Number density of the X-ray emitting material estimated from the best fits of the 10 different regions shown in Figure \ref{suzaku}.\label{density}} 
  \begin{center}
\begin{tabular}{cc}
\hline
Region & $n$ \\
 & (d$_{7}^{-0.5}$ f$^{-0.5}$ cm$^{-3}$)\\
 \hline \hline

1&$0.66_{-0.23}^{+0.28}$\\
2&$0.71_{-0.22}^{+0.24}$\\
3&$0.85_{-0.30}^{+0.31}$\\
4&$1.28_{-0.43}^{+1.17}$\\
5&$2.43_{-1.37}^{+0.96}$\\
6&$2.68_{-1.20}^{+1.01}$\\
7&$2.41_{-1.15}^{+1.28}$\\
8&$5.87_{-2.32}^{+2.82}$\\
9&$1.77_{-0.93}^{+1.32}$\\
10&$2.23_{-1.17}^{+1.41}$\\

\hline
\end{tabular}
\end{center}
\end{table}

The density of the X-ray emitting gas was calculated from the normalisation of the VVRNEI models using $n_{e} = 1.2 n_{H}$. We estimate the volume for each region by taking an area equivalent to the extracted SNR regions shown in Figure \ref{suzaku} right panel and projecting this area through a filled sphere. The estimated density $n \approx 1.1 n_{H}$ (assuming ISM abundances) is listed for each region in Table \ref{density}.
	
The inferred density ranges from $n =(0.66	_{-	0.23	}^{+	0.28	} - 5.87	_{-	2.32	}^{+	2.82	} $) d$_{7}^{-0.5}$f$^{-0.5}$ cm$^{-3}$ and is highest in region 8 which is coincident with the location of the dense molecular cloud found towards the west. The density is lowest towards the center of the remnant where the brightest X-ray emission is located, while the eastern part of the remnant has a density that is intermediate of the central regions of MSH~11$-$6{\sl 1}A. Our density estimates for the bulk of the remnant are consistent with the densities derived by \citet{2002ApJ...564..284S} who attempted to reproduce the observed temperature and brightness profiles of the remnant using the cloudy ISM model by \citet{1991ApJ...373..543W} and a hydrodynamical model that traces the evolution of the remnant, while incorporating the effects of thermal conduction.

\section{The origin of the thermal X-ray emission}

The total X-ray emitting mass in MSH~11$-$6{\sl 1}A is given by $M_{X} = 1.4 n_{H} m_{H} f V$, where $m_H$ is the mass of the hydrogen atom, $V$ is the volume from which the emission is observed, and $f$ is the filling factor. Using the estimated volumes and derived densities for regions 1-7, we sum the masses to obtain $ M_{X} \approx 480$ d$^{5/2}$f$^{1/2}$ M$_{\odot}$. This is comparable to the swept up mass derived by \citet{2002ApJ...564..284S}.

The enhancement of Mg, Si and S abundances throughout the remnant suggests that the observed X-ray emission arises in part from supernova ejecta. Assuming that all ejecta have been shocked, we can estimate the mass of the ejecta components based upon the measured abundances: $M_{i}=[(a_{i}-1)/1.4](n_{i}/n_{H})(m_{i}/m_{H})M_{X}$ where $M_{i}$ is the ejecta mass of species $i$, $a_{i}$ is its abundance relative to ISM abundances, as listed in Table 2, $m_{i}$ is the atomic mass, and $n_{i}/n_{H}$ is its ISM abundance relative to hydrogen. We find that, using the average of the measured abundances, the total ejecta masses of Mg, Si, and S are, respectively, $0.37 d^{5/2}f^{1/2}M_{\odot}$, $0.80d^{5/2}f^{1/2}M_{\odot}$, and $0.26 d^{5/2}f^{1/2}M_{\odot}$. However, we note that the abundances for Ne and Fe are both lower than ISM values, meaning that we have no evidence for ejecta components for these species, and suggesting caution in interpreting all of the abundances.  Taken at face value, however, the Mg, Si, and S ejecta mass estimates are consistent with a progenitor mass $> 25M_{\odot}$ \citep{1996ApJ...460..408T}.

Recombining plasma can arise from two main scenarios: thermal conduction which is the rapid cooling of electrons due to the interaction of the hot ejecta with the cold, dense surrounding environment \citep{1999ApJ...524..179C}; or adiabatic expansion which can occur when the SNR shockwave expands through a dense circumstellar medium into a low density ISM \citep{1989MNRAS.236..885I}. 

For thermal conduction to be the more likely scenario, the recombining plasma is expected to be coincident with the location of the molecular cloud, there should be a temperature decrease towards the molecular cloud and one would expect the thermal conduction timescale $t_{cond}$ to be less than or comparable to the age of the remnant. We find recombining plasma in regions that are directly interacting with the molecular cloud (Regions 7 and 8), and we do see a slight temperature decrease towards the molecular cloud based on the annulus regions. The thermal conduction timescale is given by \citep{1962pfig.book.....S, 2014ApJ...791...87Z} $t_{cond} \sim  k n_{e} l_{T}^{2}/\kappa \sim 56(n_{e}/1 cm^{-3})(l_{T}/10pc)^2(kT/0.6 keV)^{-5/2}(ln \Lambda/32)$kyr, where $n_{e}$ is the electron density and is calculated from our best-fits listed in Table \ref{fits}, $l_{T}$ is the scale length of the temperature gradient, $k$ is Boltzmann's constant, $\kappa$ is the thermal conductivity for a hydrogen plasma and ln $\Lambda$ is the Coulomb logarithm. Assuming a distance of 7 kpc to the remnant and a radius of $\sim 5.77'$, we calculate the length of the temperature gradient to be $\sim 3.6\times10^{19}$cm. Using the temperature and density of region 7 (see Table \ref{spectrum} and Table \ref{density} respectively), the thermal conduction timescale is estimated to be $\sim 334$ kyr. This is $\sim$16 times greater than the age of the remnant ($\sim 10-20$ kyr), making it unlikely that the overionised plasma arises via thermal conduction.

Another possibility is that the recombining plasma arises from adiabatic cooling. To calculate $t_{recomb}$, we use the best fit ionisation timescale for region 1 - 8 listed in Table \ref{fits} and divide these by the electron density of each region. We obtain a recombining timescale between $\sim(2-125)$ d$_{7}^{-0.5}$f$^{-0.5}$ kyr, which is comparable to the age of MSH~11$-$6{\sl 1}A, making this scenario the most likely. This is consistent with the results reported by \cite{2014arXiv1411.6809K} and the velocity implied by our upper limit for $kT_{init}$.

\section{The nature of $\gamma$-ray emission }


\subsection{Pulsar contribution and the Integral source ICG J11014-6103}

Pulsars that are detected within the \textit{Fermi-LAT} energy band (see the second \textit{Fermi-LAT} Pulsar catalogue by \citealt{2013ApJS..208...17A}), have spectra that is well characterised by a power law with an exponential cut-off of 1 - 5 GeV. As the $\gamma$-ray spectrum of MSH~11$-$6{\sl 1}A can be described using an exponential cut-off of $E_{cut} \sim 4.2$GeV, we still need to consider the scenario that the $\gamma$-ray emission we observe arises from a nearby pulsar other than J1105-6107. 
  
Using the Australian Telescope National Facility (ATNF) Pulsar Catalogue \citep{2005AJ....129.1993M}, there are 9 pulsars including J1105-6107 within 5$^{\circ}$, whose spin down power is sufficient to produce the $\gamma$-ray flux of MSH~11$-$6{\sl 1}A. All of these pulsars, except for J1105-6107, are >1$^{\circ}$ from the centroid of the $\gamma$-ray emission making it unlikely that any of these pulsars are contributing significantly to the observed emission of MSH~11$-$6{\sl 1}A. As we removed the contribution of J1105-6107 from the $\gamma$-ray data as described in Section 2.1, we can also rule out its contribution.
  
Recently, \citet{2014A&A...562A.122P} investigated the nature of the X-ray and radio emission of the INTEGRAL source ICG J11014-6103, which they call the lighthouse nebula. In X-rays this nebula exhibits a prominent jet-like feature that is perpendicular to an elongated cometary tail, and a point source. The source of this X-ray structure is a neutron star travelling supersonically and we have plotted its position as the white $X$ shown in Figure \ref{countmap}. This neutron star has a spin down power of $\dot{E} \sim 10^{37}$ erg s$^{-1}$. As the $\gamma$-ray luminosity of MSH~11$-$6{\sl 1}A is $2.5\times10^{35}$ erg s$^{-1}$, the PSR of ICG J11014-6103 would require an efficiency of $L_{\gamma}/ \dot{E} \sim 2.5$\% to produce the observed $\gamma$-rays, which is plausible. In an attempt to disentangle the likely source of the $\gamma$-ray emission, we have plotted as the magenta contours in Figure \ref{countmap} right panel, the \textsc{Hi} contours of the molecular cloud associated with MSH~11$-$6{\sl 1}A.  If one assumes that the pulsar of ICG J11014-6103 can produce significant $\gamma$-ray emission, we would expect the detection significance peak to be skewed towards the position of ICG J11014-6103 instead of in the direction of the remnant and MC as is observed. Thus even though we cannot rule out that ICG J11014-6103 is contributing to the $\gamma$-ray emission in Figure \ref{gammaspectrum}, the association of the molecular cloud and the detection significance in this region suggests that the emission most likely arises from the interaction of the SNR with the molecular cloud, rather than ICG J11014-6103.

\begin{table*}[t!]
\begin{center}
\caption{Model parameters and density estimates for the pion decay and leptonic model for MSH 11-61A. \label{model}}
\begin{tabular}{cccccccccccc}
\hline
Object & Distance &$\alpha_{proton}$ &$\alpha_{elec}$ &$E_{0}^{proton}$&$E_{0}^{elec}$ &Magnetic field& Ambient density & X-ray density  \\
            &(kpc)       &&  & (GeV) & (GeV) &($\mu G$)	& $n_{0}$ (cm$^{-3}$) 	& $n$ (cm$^{-3}$) 	 \\
\hline
MSH 11-61A	  &7.00& 4.39 & 3.15& 6.05&6.05 &28 &9.20 & see Table \ref{density}\\
\hline
\end{tabular}
\end{center}
\end{table*}

\subsection{Modelling the broadband emission of MSH~11$-$6{\sl 1}A}

To investigate the nature of the broadband emission from MSH~11$-$6{\sl 1}A we use a model that calculates the non-thermal emission produced by a distribution of electrons and protons. The $\pi^{0}$ decay model is based on the proton-proton interaction model by \citet{2006ApJ...647..692K}, with a scaling factor of 1.85 for helium and heavy nuclei as suggested by \citet{2009APh....31..341M}. The leptonic emission models are based on those presented by \citet{1999ApJ...513..311B} and \citet{2000ApJ...538..203B} for the synchrotron/IC and non-thermal bremsstrahlung emission mechanisms. We assume a spectral distribution of our accelerated particles to be:
\begin{equation}\label{eq:uno}
\frac{dN_{i}}{dp} = a_{i} \,p^{-\alpha_{i}} \exp\left(-\frac{p}{p_{0\,i}}\right),
\end{equation}
where $i$ is the particle species, $a_{i}$ is the normalisation of the particle distribution, $\alpha_{i}$ is the particle distribution index which is equal to (1-$\Gamma$)/2, where $\Gamma$ is the photon index and $p_{0\,i}$ is the exponential particle momentum cut-off. This distribution is transformed from momentum space to energy space such that the exponential cut-off is defined by an energy input, $E_{0i}$. The sum of the integrals of each spectral distribution is set to equal the total energy in accelerated particles within the SNR shell, $E_{CR} = \epsilon E_{SNR}$, where $\epsilon$ is the efficiency of the SNR in depositing energy into cosmic rays.

\subsubsection{Hadronic origin of the observed $\gamma$-rays}

\begin{figure*}[t!]
\begin{center}
\includegraphics[width=0.85\textwidth]{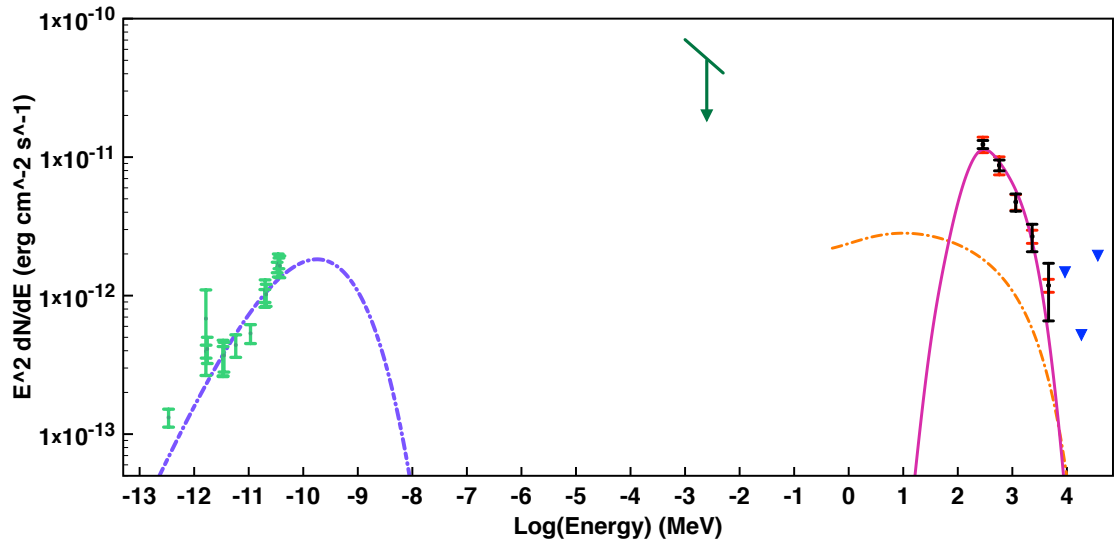}
\caption{Broadband fits to the non-thermal radio emission (green data points), non-thermal X-ray upper limit derived from the \textit{Suzaku} data (dark green limit) and the $Fermi$-LAT $\gamma$-ray emission (as described in Figure \ref{gammaspectrum}) of MSH~11$-$6{\sl 1}A. The pion decay, non-thermal bremsstrahlung and synchrotron models defined by the parameters in Table \ref{model} are shown as the solid magenta line, dashed orange line and dot-dashed purple line, respectively. The corresponding IC model falls below the plotted axis. \label{broad}} 
\end{center}
\end{figure*}

In Figure \ref{broad}, we have plotted the model fits to the broadband emission of MSH~11$-$6{\sl 1}A. The radio spectrum is a combination of multiple observations by \citet{1989PASAu...8..187M}, \cite{1996A&AS..118..329W}, and \cite{2005SerAJ.170...47F}. The X-ray upper limit was derived by fitting a power law with a photon index similar to that of RX J1713.7-3946 \citep{2003A&A...400..567U}, Kepler and RCW 86 \citep{2005ApJ...621..793B}  ($\Gamma = 2.3$), to our models of the \textit{Suzaku} data. The upper limit corresponds to the flux in which the additional non-thermal component begins to affect our reduced $\chi^{2}$. The solid magenta line corresponds to the $\pi^{0}$-decay model that adequately reproduces the observed $\gamma$-ray spectrum of MSH~11$-$6{\sl 1}A. We have also plotted as the purple dot-dashed line the synchrotron model that sufficiently reproduces the radio spectrum, assuming an electron-proton ratio ($K_{ep}$) of 0.01, while the IC model falls below the plotted axis. For completeness we have also plotted the non-thermal bremsstrahlung contribution as the orange dashed line. Table \ref{model} lists the parameters which reproduce the $\pi^{0}$-decay, synchrotron emission, IC and non-thermal bremsstrahlung models plotted in Figure \ref{broad}.
  
A $\pi^{0}$-decay model arising from a proton distribution with a power law index of $\alpha_{p}=4.34$ and a cut-off energy of $E_{0}^{p} \sim 6.05$ GeV, adequately described the $\gamma$-ray spectrum of MSH~11$-$6{\sl 1}A. The cut-off energy of the proton spectrum derived in this model is much smaller than the TeV cut-off one would expect for protons \citep{2008ARA&A..46...89R}. This could indicate that due to the high density of the surrounding environment, efficient CR acceleration is suppressed allowing accelerated particles to escape the emission volume \citep{2011NatCo...2E.194M, 2012PhPl...19h2901M}.

As non-thermal X-ray emission has not been observed from MSH~11$-$6{\sl 1}A, we are unable to constrain the cut off energy of the electron population. Thus to model the radio emission of MSH~11$-$6{\sl 1}A we assume the electron distribution has the same cut-off energy as the proton distribution. We are able to reproduce the radio spectrum using an electron distribution that has a power law index of $\alpha_{e}$=3.15 and a magnetic field of 28$\mu$G. The magnetic field implied by the synchrotron modelling is larger than the magnetic field of the ISM ($\sim 3-5\mu$G). This enhancement could arise from magnetic field amplification due to the compression of the surrounding medium by the SNR shock-wave.
  
In the $\pi^{0}$-decay model, the calculated $\gamma$-ray flux is proportional to the ambient density of the surrounding ambient medium and the total proton energy. Assuming a conservative upper limit of 40\% of the total supernova explosion energy goes into accelerating CRs, we can estimate the density of the $\gamma$-ray emitting material. For our $\pi^{0}$-decay model of MSH~11$-$6{\sl 1}A we obtain an ambient density of 9.15 cm$^{-3}$. Similar to many other SNRs that exhibit hadronic emission (e.g. W41, MSH 17-39, G337.7-0.1: \citet{2013ApJ...774...36C}; Kes 79: \citet{2014ApJ...783...32A}), this density is much larger than the ambient density estimate derived from our X-ray analysis (see Table \ref{density}). This discrepancy could arise from the SNR shock-wave interacting with dense clumps of material that are cold enough such that do not radiate significantly in X-rays \citep{2010ApJ...717..372C, 2012ApJ...744...71I}. If these clumps have a high filling factor, then the densities that we derive in our X-ray analysis would underestimate the mean local density. Our inferred ambient densities as well as the association of MSH~11$-$6{\sl 1}A with a molecular cloud towards the west of the remnant supports the conclusion that MSH~11$-$6{\sl 1}A is interacting with dense material that does not radiate  in X-rays. 
  
An alternative scenario is that the enhanced $\gamma$-ray emission arises from highly energetic particles escaping the acceleration region and are interacting with dense gas upstream of the shock (e.g. \citet{1996A&A...309..917A}, \citet{2008AIPC.1085..265G}, \citet{2008ApJ...686..325L} and \citet{2009ApJ...707L.179F}). However, a majority of these particles come from the high-energy portion of the $\gamma$-ray spectrum and the observation of ``low'' energy $\gamma$-rays may lead to inconsistencies with this scenario.

\subsubsection{Leptonic emission of the observed $\gamma$-rays}

For inverse Compton scattering to be the dominant mechanism producing the $\gamma$-rays of MSH~11$-$6{\sl 1}A, we would require greater than the entire kinetic explosion energy just in electrons, assuming that the electron to proton ratio is similar to that measured at Earth ($K_{ep}\sim0.01$) and that this emission arises from a non-thermal population of electrons being accelerated by the shock-front. This makes it difficult to conclude that IC scattering is the dominant mechanism producing the observed $\gamma$-rays. 

For non-thermal bremsstrahlung to dominate the GeV emission we require a $K_{ep} > 0.2$, assuming the maximum density derived from our X-ray analysis (see Table \ref{density}). Local measurements imply $K_{ep} \sim 0.01$ \citep{1998ApJ...492..219G}, while $\gamma$-ray modelling of other SNRs imply $K_{ep} < 0.01$ (e.g. \citealp{2010ApJ...712..287E}), making it unlikely that non-thermal bremsstrahlung emission is the dominant emission mechanism.

\section{CONCLUSION}

70 months of \textit{Fermi-LAT} data reveal significant ($\sim 5\sigma$) $\gamma$-ray emission from SNR MSH~11$-$6{\sl 1}A. This emission is consistent with being located along or inside the western limb of the remnant given the angular resolution of the \textit{Fermi-LAT} and is adjacent to regions that show a strong recombining plasma component. By modelling the broadband spectrum, we find that the emission is best described by a hadronic model, while a leptonic scenario is energetically unfavourable. This is consistent with CO and \textsc{Hi} observations that indicate the SNR is interacting with a molecular cloud towards the north and southwest. Similar to previous studies, the inferred density from our pion decay model is much higher than that implied by the thermal X-ray emission. $Suzaku$ data reveal that the bulk of the X-ray emission of MSH~11$-$6{\sl 1}A arises from a single recombining plasma with enhanced abundances of Mg, Si and S with some regions also requiring an underabundance of Ne and Fe, while the emission towards the east of the remnant arises from an ionising plasma with Mg, Si and S. The origin of the recombining plasma is most like adiabatic cooling. We find that the results from our central regions (1 - 3) and our regions 9 - 10, agree well with those that \citet{2014arXiv1411.6809K} obtained for their corresponding regions. The enhancement of Mg, Si and S suggests that some of the observed emission arises from shocked ejecta and that the progenitor of MSH~11$-$6{\sl 1}A had a mass $>25M_{\odot}$.

\bibliography{msh1161a_accepted}

\begin{thebibliography}{}
\expandafter\ifx\csname natexlab\endcsname\relax\def\natexlab#1{#1}\fi

\bibitem[{{Abdo} {et~al.}(2010{\natexlab{a}}){Abdo}, Ackermann, {Ajello},
  {et~al.}}]{2010ApJ...722.1303A}
{Abdo}, A.~A., Ackermann, M., {Ajello}, M., {et~al.} 2010{\natexlab{a}}, \apj,
  722, 1303

\bibitem[{{Abdo} {et~al.}(2010{\natexlab{b}}){Abdo}, {Ackermann}, {Ajello},
  {et~al.}}]{2010Sci...327.1103A}
{Abdo}, A.~A., {Ackermann}, M., {Ajello}, M., {et~al.} 2010{\natexlab{b}},
  Science, 327, 1103

\bibitem[{{Abdo} {et~al.}(2013){Abdo}, {Ajello}, {Allafort},
  {et~al.}}]{2013ApJS..208...17A}
{Abdo}, A.~A., {Ajello}, M., {Allafort}, A., {et~al.} 2013, \apjs, 208, 17

\bibitem[{Ackermann {et~al.}(2012)Ackermann, Ajello, Albert,
  {et~al.}}]{2012ApJS..203....4A}
Ackermann, M., Ajello, M., Albert, A., {et~al.} 2012, \apjs, 203, 4

\bibitem[{{Ackermann} {et~al.}(2013){Ackermann}, Ajello, Allafort,
  {et~al.}}]{2013Sci...339..807A}
{Ackermann}, M., Ajello, M., Allafort, A., {et~al.} 2013, Science, 339, 807

\bibitem[{{Aharonian} \& {Atoyan}(1996)}]{1996A&A...309..917A}
{Aharonian}, F.~A., \& {Atoyan}, A.~M. 1996, \aap, 309, 917

\bibitem[{{Anders} \& {Grevesse}(1989)}]{1989GeCoA..53..197A}
{Anders}, E., \& {Grevesse}, N. 1989, GeCoA, 53, 197

\bibitem[{{Auchettl} {et~al.}(2014){Auchettl}, {Slane}, \&
  {Castro}}]{2014ApJ...783...32A}
{Auchettl}, K., {Slane}, P., \& {Castro}, D. 2014, \apj, 783, 32

\bibitem[{{Bamba} {et~al.}(2005){Bamba}, {Yamazaki}, {Yoshida},
  {et~al.}}]{2005ApJ...621..793B}
{Bamba}, A., {Yamazaki}, R., {Yoshida}, T., {et~al.} 2005, \apj, 621, 793

\bibitem[{{Baring} {et~al.}(1999){Baring}, {Ellison}, {Reynolds},
  {et~al.}}]{1999ApJ...513..311B}
{Baring}, M.~G., {Ellison}, D.~C., {Reynolds}, S.~P., {et~al.} 1999, \apj, 513,
  311

\bibitem[{{Brand} \& {Blitz}(1993)}]{1993A&A...275...67B}
{Brand}, J., \& {Blitz}, L. 1993, \aap, 275, 67

\bibitem[{{Bykov} {et~al.}(2000){Bykov}, {Chevalier}, {Ellison},
  {et~al.}}]{2000ApJ...538..203B}
{Bykov}, A.~M., {Chevalier}, R.~A., {Ellison}, D.~C., {et~al.} 2000, \apj, 538,
  203

\bibitem[{{Caraveo}(1993)}]{1993ApJ...415L.111C}
{Caraveo}, P.~A. 1993, \apjl, 415, L111

\bibitem[{{Castro} \& {Slane}(2010)}]{2010ApJ...717..372C}
{Castro}, D., \& {Slane}, P. 2010, \apj, 717, 372

\bibitem[{{Castro} {et~al.}(2013){Castro}, {Slane}, {Carlton},
  {et~al.}}]{2013ApJ...774...36C}
{Castro}, D., {Slane}, P., {Carlton}, A., {et~al.} 2013, \apj, 774, 36

\bibitem[{{Cordes} \& {Lazio}(2002)}]{2002astro.ph..7156C}
{Cordes}, J.~M., \& {Lazio}, T.~J.~W. 2002, astro-ph/0207156

\bibitem[{{Cox} {et~al.}(1999){Cox}, {Shelton}, {Maciejewski},
  {et~al.}}]{1999ApJ...524..179C}
{Cox}, D.~P., {Shelton}, R.~L., {Maciejewski}, W., {et~al.} 1999, \apj, 524,
  179

\bibitem[{{Dickel}(1973)}]{1973ApL....15...61D}
{Dickel}, J.~R. 1973, ApL, 15, 61

\bibitem[{{Dwarkadas}(2005)}]{2005ApJ...630..892D}
{Dwarkadas}, V.~V. 2005, \apj, 630, 892

\bibitem[{{Ellison} {et~al.}(2010){Ellison}, {Patnaude}, {Slane},
  {et~al.}}]{2010ApJ...712..287E}
{Ellison}, D.~C., {Patnaude}, D.~J., {Slane}, P., {et~al.} 2010, \apj, 712, 287

\bibitem[{{Ergin} {et~al.}(2014){Ergin}, {Sezer}, {Saha},
  {et~al.}}]{2014ApJ...790...65E}
{Ergin}, T., {Sezer}, A., {Saha}, L., {et~al.} 2014, \apj, 790, 65

\bibitem[{{Filipovic} {et~al.}(2005){Filipovic}, {Payne}, \&
  {Jones}}]{2005SerAJ.170...47F}
{Filipovic}, M.~D., {Payne}, J.~L., \& {Jones}, P.~A. 2005, SerAJ, 170, 47

\bibitem[{{Foster} {et~al.}(2012){Foster}, {Ji}, {Smith},
  {et~al.}}]{2012ApJ...756..128F}
{Foster}, A.~R., {Ji}, L., {Smith}, R.~K., {et~al.} 2012, \apj, 756, 128

\bibitem[{{Fujita} {et~al.}(2009){Fujita}, {Ohira}, {Tanaka},
  {et~al.}}]{2009ApJ...707L.179F}
{Fujita}, Y., {Ohira}, Y., {Tanaka}, S.~J., {et~al.} 2009, \apjl, 707, L179

\bibitem[{{Gabici} {et~al.}(2008){Gabici}, {Casanova}, \&
  {Aharonian}}]{2008AIPC.1085..265G}
{Gabici}, S., {Casanova}, S., \& {Aharonian}, F.~A. 2008, AIP Conf. Proc.,
  1085, 265

\bibitem[{{Gaisser} {et~al.}(1998){Gaisser}, {Protheroe}, \&
  {Stanev}}]{1998ApJ...492..219G}
{Gaisser}, T.~K., {Protheroe}, R.~J., \& {Stanev}, T. 1998, \apj, 492, 219

\bibitem[{{Garc{\'{\i}}a} {et~al.}(2012){Garc{\'{\i}}a}, {Combi},
  {Albacete-Colombo}, {et~al.}}]{2012A&A...546A..91G}
{Garc{\'{\i}}a}, F., {Combi}, J.~A., {Albacete-Colombo}, J.~F., {et~al.} 2012,
  \aap, 546, A91

\bibitem[{{Goss} {et~al.}(1972){Goss}, {Radhakrishnan}, {Brooks},
  {et~al.}}]{1972ApJS...24..123G}
{Goss}, W.~M., {Radhakrishnan}, V., {Brooks}, J.~W., {et~al.} 1972, \apjs, 24,
  123

\bibitem[{{Henley} \& {Shelton}(2013)}]{2013ApJ...773...92H}
{Henley}, D.~B., \& {Shelton}, R.~L. 2013, \apj, 773, 92

\bibitem[{{Hobbs} {et~al.}(2005){Hobbs}, {Lorimer}, {Lyne},
  {et~al.}}]{2005MNRAS.360..974H}
{Hobbs}, G., {Lorimer}, D.~R., {Lyne}, A.~G., {et~al.} 2005, \mnras, 360, 974

\bibitem[{{Inoue} {et~al.}(2012){Inoue}, {Yamazaki}, {Inutsuka},
  {et~al.}}]{2012ApJ...744...71I}
{Inoue}, T., {Yamazaki}, R., {Inutsuka}, S., {et~al.} 2012, \apj, 744, 71

\bibitem[{{Itoh} \& {Masai}(1989)}]{1989MNRAS.236..885I}
{Itoh}, H., \& {Masai}, K. 1989, \mnras, 236, 885

\bibitem[{{Kamae} {et~al.}(2006){Kamae}, {Karlsson}, {Mizuno},
  {et~al.}}]{2006ApJ...647..692K}
{Kamae}, T., {Karlsson}, N., {Mizuno}, T., {et~al.} 2006, \apj, 647, 692

\bibitem[{{Kamitsukasa} {et~al.}(2015){Kamitsukasa}, {Koyama}, {Uchida},
  {et~al.}}]{2014arXiv1411.6809K}
{Kamitsukasa}, F., {Koyama}, K., {Uchida}, H., {et~al.} 2015, \pasj, 67, 16

\bibitem[{{Kaneda} {et~al.}(1997){Kaneda}, {Makishima}, {Yamauchi},
  {et~al.}}]{1997ApJ...491..638K}
{Kaneda}, H., {Makishima}, K., {Yamauchi}, S., {et~al.} 1997, \apj, 491, 638

\bibitem[{{Kaspi} {et~al.}(1997){Kaspi}, {Bailes}, {Manchester},
  {et~al.}}]{1997ApJ...485..820K}
{Kaspi}, V.~M., {Bailes}, M., {Manchester}, R.~N., {et~al.} 1997, \apj, 485,
  820

\bibitem[{{Kaspi} {et~al.}(1996){Kaspi}, {Manchester}, {Johnston},
  {et~al.}}]{1996AJ....111.2028K}
{Kaspi}, V.~M., {Manchester}, R.~N., {Johnston}, S., {et~al.} 1996, \aj, 111,
  2028

\bibitem[{{Kawasaki} {et~al.}(2005){Kawasaki}, {Ozaki}, {Nagase},
  {et~al.}}]{2005ApJ...631..935K}
{Kawasaki}, M., {Ozaki}, M., {Nagase}, F., {et~al.} 2005, \apj, 631, 935

\bibitem[{{Kawasaki} {et~al.}(2002){Kawasaki}, {Ozaki}, {Nagase},
  {et~al.}}]{2002ApJ...572..897K}
{Kawasaki}, M.~T., {Ozaki}, M., {Nagase}, F., {et~al.} 2002, \apj, 572, 897

\bibitem[{{Kesteven} \& {Caswell}(1987)}]{1987A&A...183..118K}
{Kesteven}, M.~J., \& {Caswell}, J.~L. 1987, \aap, 183, 118

\bibitem[{{Kesteven}(1968)}]{1968AuJPh..21..369K}
{Kesteven}, M.~J.~L. 1968, AuJPh, 21, 369

\bibitem[{{Koyama} {et~al.}(2007){Koyama}, {Tsunemi}, {Dotani},
  {et~al.}}]{2007PASJ...59S..23K}
{Koyama}, K., {Tsunemi}, H., {Dotani}, T., {et~al.} 2007, \pasj, 59, 23

\bibitem[{{Kramer} {et~al.}(2003){Kramer}, {Bell}, {Manchester},
  {et~al.}}]{2003MNRAS.342.1299K}
{Kramer}, M., {Bell}, J.~F., {Manchester}, R.~N., {et~al.} 2003, \mnras, 342,
  1299

\bibitem[{{Kushino} {et~al.}(2002){Kushino}, {Ishisaki}, {Morita},
  {et~al.}}]{2002PASJ...54..327K}
{Kushino}, A., {Ishisaki}, Y., {Morita}, U., {et~al.} 2002, \pasj, 54, 327

\bibitem[{{Lee} {et~al.}(2008){Lee}, {Kamae}, \&
  {Ellison}}]{2008ApJ...686..325L}
{Lee}, S.-H., {Kamae}, T., \& {Ellison}, D.~C. 2008, \apj, 686, 325

\bibitem[{{Malkov} {et~al.}(2011){Malkov}, {Diamond}, \&
  {Sagdeev}}]{2011NatCo...2E.194M}
{Malkov}, M.~A., {Diamond}, P.~H., \& {Sagdeev}, R.~Z. 2011, NatCo, 2, 194

\bibitem[{{Malkov} {et~al.}(2012){Malkov}, {Diamond}, \&
  {Sagdeev}}]{2012PhPl...19h2901M}
---. 2012, PhPl, 19, 082901

\bibitem[{{Manchester} {et~al.}(2005){Manchester}, {Hobbs}, {Teoh},
  {et~al.}}]{2005AJ....129.1993M}
{Manchester}, R.~N., {Hobbs}, G.~B., {Teoh}, A., {et~al.} 2005, \aj, 129, 1993

\bibitem[{{McClure-Griffiths} {et~al.}(2005){McClure-Griffiths}, {Dickey},
  {Gaensler}, {et~al.}}]{2005ApJS..158..178M}
{McClure-Griffiths}, N.~M., {Dickey}, J.~M., {Gaensler}, B.~M., {et~al.} 2005,
  \apjs, 158, 178

\bibitem[{{Mills} {et~al.}(1961){Mills}, {Slee}, \&
  {Hill}}]{1961AuJPh..14..497M}
{Mills}, B.~Y., {Slee}, O.~B., \& {Hill}, E.~R. 1961, AuJPh, 14, 497

\bibitem[{{Milne} {et~al.}(1989){Milne}, {Caswell}, {Kesteven},
  {et~al.}}]{1989PASAu...8..187M}
{Milne}, D.~K., {Caswell}, J.~L., {Kesteven}, M.~J., {et~al.} 1989, PASAu, 8,
  187

\bibitem[{{Mori}(2009)}]{2009APh....31..341M}
{Mori}, M. 2009, APh, 31, 341

\bibitem[{{Moriya}(2012)}]{2012ApJ...750L..13M}
{Moriya}, T.~J. 2012, \apjl, 750, L13

\bibitem[{{Nolan} {et~al.}(2012){Nolan}, {Abdo}, {Ackermann},
  {et~al.}}]{2012ApJS..199...31N}
{Nolan}, P.~L., {Abdo}, A.~A., {Ackermann}, M., {et~al.} 2012, \apjs, 199, 31

\bibitem[{{Pavan} {et~al.}(2014){Pavan}, {Bordas}, {P{\"u}hlhofer},
  {et~al.}}]{2014A&A...562A.122P}
{Pavan}, L., {Bordas}, P., {P{\"u}hlhofer}, G., {et~al.} 2014, \aap, 562, A122

\bibitem[{{Reynolds}(2008)}]{2008ARA&A..46...89R}
{Reynolds}, S.~P. 2008, \araa, 46, 89

\bibitem[{{Reynoso} {et~al.}(2006){Reynoso}, {Johnston}, {Green},
  {et~al.}}]{2006MNRAS.369..416R}
{Reynoso}, E.~M., {Johnston}, S., {Green}, A.~J., {et~al.} 2006, \mnras, 369,
  416

\bibitem[{{Rho} \& {Petre}(1998)}]{1998ApJ...503L.167R}
{Rho}, J., \& {Petre}, R. 1998, \apjl, 503, L167

\bibitem[{{Rosado} {et~al.}(1996){Rosado}, {Ambrocio-Cruz}, {Le Coarer},
  {et~al.}}]{1996A&A...315..243R}
{Rosado}, M., {Ambrocio-Cruz}, P., {Le Coarer}, E., {et~al.} 1996, \aap, 315,
  243

\bibitem[{{Sato} {et~al.}(2014){Sato}, {Koyama}, {Takahashi},
  {et~al.}}]{2014arXiv1409.6861S}
{Sato}, T., {Koyama}, K., {Takahashi}, T., {et~al.} 2014, \pasj, 66, 124

\bibitem[{{Seward}(1990)}]{1990ApJS...73..781S}
{Seward}, F.~D. 1990, \apjs, 73, 781

\bibitem[{{Slane} {et~al.}(2015){Slane}, {Bykov}, {Ellison},
  {et~al.}}]{2014SSRv..tmp...26S}
{Slane}, P., {Bykov}, A., {Ellison}, D.~C., {et~al.} 2015, \ssr, 188, 187

\bibitem[{{Slane} {et~al.}(2002){Slane}, {Smith}, {Hughes},
  {et~al.}}]{2002ApJ...564..284S}
{Slane}, P., {Smith}, R.~K., {Hughes}, J.~P., {et~al.} 2002, \apj, 564, 284

\bibitem[{{Smith} {et~al.}(2001){Smith}, {Brickhouse}, {Liedahl},
  {et~al.}}]{2001ApJ...556L..91S}
{Smith}, R.~K., {Brickhouse}, N.~S., {Liedahl}, D.~A., {et~al.} 2001, \apjl,
  556, L91

\bibitem[{{Smith} \& {Hughes}(2010)}]{2010ApJ...718..583S}
{Smith}, R.~K., \& {Hughes}, J.~P. 2010, \apj, 718, 583

\bibitem[{{Spitzer}(1962)}]{1962pfig.book.....S}
{Spitzer}, L. 1962, Physics of Fully Ionized Gases, (2nd ed.; New York:
  Interscience)

\bibitem[{{Tawa} {et~al.}(2008){Tawa}, {Hayashida}, {Nagai},
  {et~al.}}]{2008PASJ...60S..11T}
{Tawa}, N., {Hayashida}, K., {Nagai}, M., {et~al.} 2008, \pasj, 60, 11

\bibitem[{{Thielemann} {et~al.}(1996){Thielemann}, {Nomoto}, \&
  {Hashimoto}}]{1996ApJ...460..408T}
{Thielemann}, F.-K., {Nomoto}, K., \& {Hashimoto}, M.-A. 1996, \apj, 460, 408

\bibitem[{{Uchiyama} {et~al.}(2003){Uchiyama}, {Aharonian}, \&
  {Takahashi}}]{2003A&A...400..567U}
{Uchiyama}, Y., {Aharonian}, F.~A., \& {Takahashi}, T. 2003, \aap, 400, 567

\bibitem[{{Uchiyama} {et~al.}(2007){Uchiyama}, {Aharonian}, {Tanaka},
  {et~al.}}]{2007Natur.449..576U}
{Uchiyama}, Y., {Aharonian}, F.~A., {Tanaka}, T., {et~al.} 2007, \nat, 449, 576

\bibitem[{{Warren} {et~al.}(2005){Warren}, {Hughes}, {Badenes},
  {et~al.}}]{2005ApJ...634..376W}
{Warren}, J.~S., {Hughes}, J.~P., {Badenes}, C., {et~al.} 2005, \apj, 634, 376

\bibitem[{{White} \& {Long}(1991)}]{1991ApJ...373..543W}
{White}, R.~L., \& {Long}, K.~S. 1991, \apj, 373, 543

\bibitem[{{Whiteoak} \& {Green}(1996)}]{1996A&AS..118..329W}
{Whiteoak}, J.~B.~Z., \& {Green}, A.~J. 1996, \aaps, 118, 329

\bibitem[{{Wilms} {et~al.}(2000){Wilms}, {Allen}, \&
  {McCray}}]{2000ApJ...542..914W}
{Wilms}, J., {Allen}, A., \& {McCray}, R. 2000, \apj, 542, 914

\bibitem[{{Yu} {et~al.}(2013){Yu}, {Manchester}, {Hobbs},
  {et~al.}}]{2013MNRAS.429..688Y}
{Yu}, M., {Manchester}, R.~N., {Hobbs}, G., {et~al.} 2013, \mnras, 429, 688

\bibitem[{{Zhou} {et~al.}(2014){Zhou}, {Safi-Harb}, {Chen},
  {et~al.}}]{2014ApJ...791...87Z}
{Zhou}, P., {Safi-Harb}, S., {Chen}, Y., {et~al.} 2014, \apj, 791, 87

\end{thebibliography}

\end{document}